\newcommand{\CLASSINPUTtoptextmargin}{19mm}%
\newcommand{\CLASSINPUTbottomtextmargin}{43mm}%
\newcommand{\CLASSINPUTinnersidemargin}{12.9mm}%
\newcommand{\CLASSINPUToutersidemargin}{12.9mm}%
\documentclass[conference,10pt,a4paper]{IEEEtran}
\usepackage{amsmath}
\usepackage{times}
\usepackage{graphicx}
\usepackage{multirow}
\usepackage[none]{hyphenat}
\usepackage{float}
\usepackage{subfig}
\usepackage [utf8]{inputenc}
\usepackage{gensymb}
\usepackage{enumitem}
\usepackage{psfrag}
\usepackage{xcolor}
\usepackage{units}

\usepackage[utf8]{inputenc}
\usepackage{cite}
\usepackage{amsmath,amssymb,amsfonts}
\usepackage{algorithmic}
\usepackage{graphicx}
\usepackage{textcomp}
\usepackage{hyperref}
\usepackage{color,soul}
\usepackage{units}
\usepackage{textcomp}
\usepackage[T1]{fontenc}
\usepackage{wrapfig}
\usepackage{tikz}

\def\BibTeX{{\rm B\kern-.05em{\sc i\kern-.025em b}\kern-.08em
    T\kern-.1667em\lower.7ex\hbox{E}\kern-.125emX}}
\usepackage{float}

\makeatletter

\def\@maketitle{\newpage
\bgroup\par\addvspace{0.5\baselineskip}\centering%
\ifCLASSOPTIONtechnote
   {\bfseries\large\@IEEEcompsoconly{\sffamily}\@title\par}\vskip 1.3em{\lineskip .5em\@IEEEcompsoconly{\sffamily}\@author
   \@IEEEspecialpapernotice\par{\@IEEEcompsoconly{\vskip 1.5em\relax
   \@IEEEtitleabstractindextextbox{\@IEEEtitleabstractindextext}\par
   \hfill\@IEEEcompsocdiamondline\hfill\hbox{}\par}}}\relax
\else
   \vskip0.2em{\EuMWtitlesize\ifCLASSOPTIONtransmag\bfseries\LARGE\fi\@IEEEcompsoconly{\sffamily}\@IEEEcompsocconfonly{\normalfont\normalsize\vskip 2\@IEEEnormalsizeunitybaselineskip
   \bfseries\Large}\@title\par}\vskip1.0em\par
   \ifCLASSOPTIONconference%
      {\@IEEEspecialpapernotice\mbox{}\vskip\@IEEEauthorblockconfadjspace%
       \mbox{}\hfill\begin{@IEEEauthorhalign}\@author\end{@IEEEauthorhalign}\hfill\mbox{}\par}\relax
   \else
      \ifCLASSOPTIONpeerreviewca
         {\@IEEEcompsoconly{\sffamily}\@IEEEspecialpapernotice\mbox{}\vskip\@IEEEauthorblockconfadjspace%
          \mbox{}\hfill\begin{@IEEEauthorhalign}\@author\end{@IEEEauthorhalign}\hfill\mbox{}\par
          {\@IEEEcompsoconly{\vskip 1.5em\relax
           \@IEEEtitleabstractindextextbox{\@IEEEtitleabstractindextext}\par\hfill
           \@IEEEcompsocdiamondline\hfill\hbox{}\par}}}\relax
      \else
         \ifCLASSOPTIONtransmag
           {\@IEEEspecialpapernotice\mbox{}\vskip\@IEEEauthorblockconfadjspace%
            \mbox{}\hfill\begin{@IEEEauthorhalign}\@author\end{@IEEEauthorhalign}\hfill\mbox{}\par
           {\vspace{0.5\baselineskip}\relax\@IEEEtitleabstractindextextbox{\@IEEEtitleabstractindextext}\vspace{-1\baselineskip}\par}}\relax
         \else
           {\lineskip.5em\@IEEEcompsoconly{\sffamily}\sublargesize\@author\@IEEEspecialpapernotice\par
           {\@IEEEcompsoconly{\vskip 1.5em\relax
            \@IEEEtitleabstractindextextbox{\@IEEEtitleabstractindextext}\par\hfill
            \@IEEEcompsocdiamondline\hfill\hbox{}\par}}}\relax
         \fi
      \fi
   \fi
\fi\par\addvspace{0.0\baselineskip}\egroup}

\def\EuMWtitlesize{\@setfontsize{\EuMWtitlesize}{24}{24pt}}
\def\EuMWauthorsize{\@setfontsize{\EuMWauthorsize}{11}{11pt}}
\def\EuMWaffilsize{\@setfontsize{\EuMWaffilsize}{10}{10pt}}
\def\EuMWcaptionsize{\@setfontsize{\EuMWcaptionsize}{9}{10pt}}
\def\EuMWbibsize{\@setfontsize{\EuMWbibsize}{8}{10pt}}

\def\@IEEEauthorblockNstyle{\EuMWauthorsize\@IEEEcompsocnotconfonly{\sffamily}\@IEEEcompsocconfonly{\large}}
\def\@IEEEauthorblockAstyle{\EuMWaffilsize\@IEEEcompsocnotconfonly{\sffamily}\@IEEEcompsocconfonly{\itshape}\@IEEEcompsocconfonly{\large}}
\def\@IEEEauthordefaulttextstyle{\EuMWauthorsize\@IEEEcompsocnotconfonly{\sffamily}\sublargesize}

\def\thebibliography#1{\section*{\refname}%
    \addcontentsline{toc}{section}{\refname}%
    \EuMWbibsize\@IEEEcompsocconfonly{\small}\vskip 0.3\baselineskip plus 0.1\baselineskip minus 0.1\baselineskip
    \list{\@biblabel{\@arabic\c@enumiv}}%
    {\settowidth\labelwidth{\@biblabel{#1}}%
    \leftmargin\labelwidth
    \advance\leftmargin\labelsep\relax
    \itemsep \IEEEbibitemsep\relax
    \usecounter{enumiv}%
    \let\p@enumiv\@empty
    \renewcommand\theenumiv{\@arabic\c@enumiv}}%
    \let\@IEEElatexbibitem\bibitem%
    \def\bibitem{\@IEEEbibitemprefix\@IEEElatexbibitem}%
\def\newblock{\hskip .11em plus .33em minus .07em}%
\ifCLASSOPTIONtechnote\sloppy\clubpenalty4000\widowpenalty4000\interlinepenalty100%
\else\sloppy\clubpenalty4000\widowpenalty4000\interlinepenalty500\fi%
    \sfcode`\.=1000\relax}

\long\def\@makecaption#1#2{%
\ifx\@captype\@IEEEtablestring%
\par\@IEEEtabletopskipstrut
\else
\@IEEEfigurecaptionsepspace
\fi
\setbox\@tempboxa\hbox{\normalfont\footnotesize {#1.}\nobreakspace\nobreakspace #2}%
\ifdim \wd\@tempboxa >\hsize%
\setbox\@tempboxa\hbox{\normalfont\footnotesize {#1.}\nobreakspace\nobreakspace}%
\parbox[t]{\hsize}{\normalfont\footnotesize\noindent\unhbox\@tempboxa#2}%
\else
\ifCLASSOPTIONconference \hbox to\hsize{\normalfont\footnotesize\hfil\box\@tempboxa\hfil}%
\else \hbox to\hsize{\normalfont\footnotesize\box\@tempboxa\hfil}%
\fi\fi
\ifx\@captype\@IEEEtablestring%
\@IEEEtablecaptionsepspace
\else
\fi}

\newlength\tablecaptiontotableskip
\newlength\figuretocaptionskip
\def\@IEEEfigurecaptionsepspace{\vskip\figuretocaptionskip\relax}%
\def\@IEEEtablecaptionsepspace{\vskip\tablecaptiontotableskip\relax}%

\def\abstract{\normalfont%
\@IEEEabskeysecsize\bfseries\textit{\abstractname}\,\bfseries\textit{---}\,%
\@IEEEgobbleleadPARNLSP}%

\def\IEEEkeywords{\normalfont%
\@IEEEabskeysecsize\bfseries\textit{\IEEEkeywordsname}\,\bfseries\textit{---}\,%
\@IEEEgobbleleadPARNLSP}%
\def\endIEEEkeywords{\relax\vspace{0.67ex}%
\par\if@twocolumn\else\endquotation\fi%
\normalsize\normalfont}%

\def\@IEEEauthorblockNtopspace{0ex}
\def\@IEEEauthorblockAtopspace{1mm}
\def\tablename{Table}
\def\thetable{\arabic{table}}
\def\IEEEkeywordsname{Keywords}
\def\subsubsection{\@startsection{subsubsection}{3}{\z@}{1.5ex plus 1.5ex minus 0.5ex}%
{0.7ex plus .5ex minus 0ex}{\normalfont\normalsize\itshape}}%
\newlength{\CPheadmatchindent}%
\def\@seccntformat#1{\hbox to\CPheadmatchindent{\csname the#1dis\endcsname}\hskip 0.1em \relax}
\IEEEilabelindentA \parindent
\IEEEilabelindent \IEEEilabelindentA
\IEEEelabelindent \parindent
\IEEEdlabelindent \parindent
\IEEElabelindent \parindent
\makeatother

\begin{document}
\raggedbottom


\title{Beam Divergence Reduction of Vortex Waves with a Tailored Lens and a Tailored Reflector}


\author{%
\IEEEauthorblockN{%
M. Haj Hassan$^{1}$,
B. Sievert$^{1}$,
J. T. Svejda$^{1}$,
A. Mostafa Ahmad$^{2}$,
J. Barowski$^{3}$,\\
A. Rennings$^{1}$,
I. Rolfes$^{3}$, 
A. Sezgin$^{2}$,
and D. Erni$^{1}$
}
\IEEEauthorblockA{%
\ $^{1}$General and Theoretical Electrical Engineering (ATE), Faculty of Engineering, and
CENIDE – \\ Center for Nanointegration Duisburg-Essen, University of Duisburg Essen, D-47048 Duisburg, Germany\\
\ $^{2}$Department of Electrical Engineering, Ruhr-Universität Bochum, 44801 Bochum, Germany\\
\ $^{3}$Institute of Microwave Systems Ruhr-University Bochum, Germany
}
}
%
\maketitle


\begin{abstract}
    In this paper, we present a tailored lens and a tailored reflector in order to reduce the
    large beam divergence inherent to Orbital Angular Momentum waves (OAM waves/Vortex waves) 
    that are generated by an Uniform Circular 
    Patch Antenna Array (UCA) at $\unit[10]{GHz}$. 
    The tailored lens and the tailored reflector are designed by the shape function \eqref{eq:Fermat_lens}
    and \eqref{eq:Fermat_Reflector} around the antenna's center axis, respectively.
    The tailored lens is compared to UCA without and with conventional lens.
    The simulated and measured results show a significant improvement when using the tailored lens.
    Following, a tailored reflector is implemented and compared to UCA without and with conventional reflector.
    Firstly, the two reflectors are simulated with an impressed field source to 
    neglect the influence of the UCA. 
    The simulated results of the two reflectors show that the tailored 
    reflector has a better performance than the 
    conventional reflector when the height of the reflector
    $r_0$ is less than around $\unit[1.5]{}$$\lambda$ and when the opening angle of the
    reflector $\vartheta$ is less than $\unit[38]{}$$^\circ$ (for UCA with $d$ of $\lambda/2$). 
    In addition, the reflectors are simulated with real UCA with
    different shapes of PCB,
    which can disturb the reflected waves from the reflector.
    Two lenses and two reflectors are manufactured and measured 
    in an anechoic chamber and compared to the simulation results. 
    This paper shows that the vortex waves needs a special lens 
    or a special reflector to reduce effictively the beam divergence especially 
    when the radius of the UCA is very large.
\end{abstract}



\section{Introduction}
In recent years, vortex waves have attracted the interest of 
many scientists, especially after the successful utilization of 
vortex waves in the optics domain \cite{OAm_optics}.
Electromagnetic waves can carry Spin Angular Momentum (SAM) (intrinsic rotation) 
and Orbital Angular Momentum (OAM) (extrinsic rotation i.e 
macroscopic helical phase) \cite{intrinsic}.
The vortex waves are characterized by a 
helical phase distribution that changes linearly around the beam axis, a doughnut-shaped radiation pattern, 
and by the phase singularity, i.e. the phase along the 
beam axis is not determinable. The following equation describes the 
vortex waves $\mathbf{E}(\rho,\varphi)= E_0(\rho) \mathrm{exp}(j\varphi m)\mathbf{e}_r$, 
where $E_0(\rho)$ is the amplitude of the electric field strength,
$\varphi$ is the geometric azimuthal angle and $m$ is the OAM mode order. 
There are many possibilities to generate OAM waves such 
as elliptical patch antennas \cite{ellipticalpatch}, uniform 
circular patch antenna arrays (UCA) \cite{patchantennaarray}, 
spiral phase plates \cite{SPP}, holographic plates \cite{HP}, 
metasurfaces \cite{metasurface}, and  reflectors \cite{Reflector}. 
Each of these approaches have advantages and disadvantages, 
regarding e.g. costs, fabrication simplicity, integration capability, 
simple design and implementation, simple feeding etc.
Many researchers have published regarding OAM beams with different mode orders 
$m$ ($...,-2,-1,0,1,2,...$), which is an additional degree of freedom 
in signal coding. This feature of the OAM yields additional capacity and spectral efficiency enabling the 
transmission of multiple signals of the same frequency 
within the same time interval.
Nevertheless, the vortex waves are suffering from the increased beam divergence,
especially when higher OAM mode orders are utilized \cite{Beamdivergence}.
This large beam divergence can be an issue for many applications,
such as in the wireless communication.
Several publications tried to reduce the beam divergence 
by using Fabry-Perot Cavity \cite{Fabryperot}, by using lenses 
\cite{Lens1}, \cite{Lens2}, \cite{Lens3}, \cite{Lens4},
by using antenna arrays in UCA \cite{ArrayUCA}
, and by using reflectors \cite{Reflector1}, \cite{Reflector2}, \cite{Reflector3} ,
\cite{Reflector4}, \cite{Reflector5}.
Each of these approaches have issues with the design, with the integration, or with the manufacturing.
Here, we utilize the uniform circular patch antenna array approach and design
a tailored lens and a tailored reflector to overcome the inherent divergence 
of the vortex beams. The simulated and measured results of the two approaches 
are compared to UCA and to the conventional lens and conventional reflector.
In the following, we review the simulated and the measured results of the
lens in Section II and III. Then, we show the design process of the reflector
with impressed source and with real UCA in Section IV and V. In Section VI, we present 
the measured results of the reflector.


\section{Design of Uniform Circular Array (UCA)}

\begin{figure}[H]
\centering
\includegraphics[width=83mm]{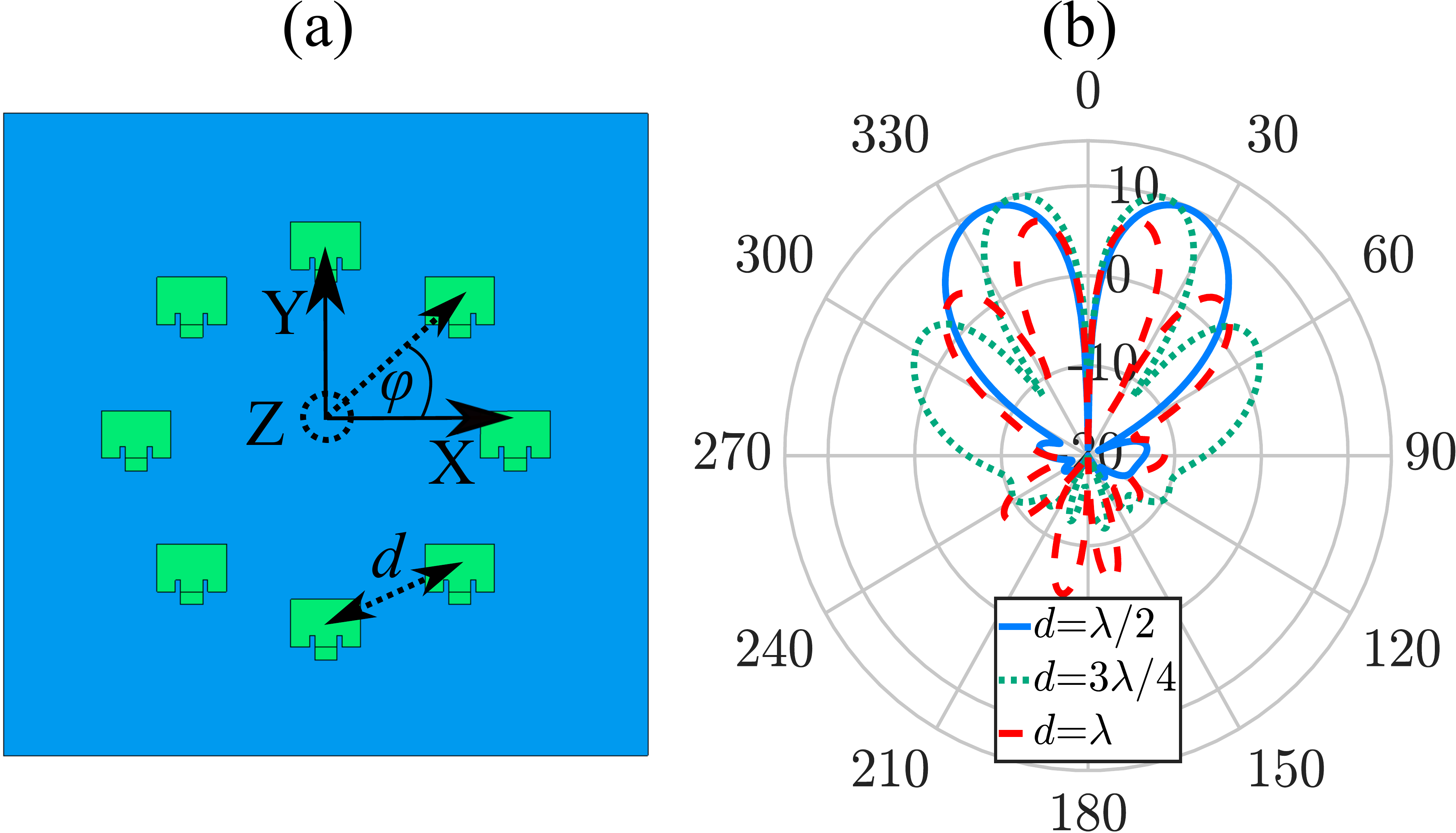}
\caption{Top view of the UCA (a), radiation pattern of the UCA with the mode order $-1$ at $\varphi  = 0^\circ$ for an 
element separation $d$ of $\lambda/2$, 3$\lambda/4$, and $\lambda$ 
($\lambda_0=\unit[30]{mm}$ at $\unit[10]{GHz}$) (b).}
\label{fig:UCA}
\end{figure}

At first, a single rectangular patch antenna element is designed 
on a $\unit[30]{mm} \times \unit[30]{mm}$ printed circuit board (PCB). 
The length and the width of the patch 
antenna are $\unit[7.4]{mm}$ (about $\lambda{_\mathrm{eff}/2}$) and $\unit[10.8]{mm}$, 
respectively. Two insets in the antenna enable the matching 
of the antenna to $\unit[50]{}$ $\Omega$ in order to maximize the realized gain, 
which is $\unit[7]{dBi}$. The antenna has a
Y-polarization with respect to the coordinate system given in Fig. \ref{fig:UCA}. In the following, this single 
antenna is extended to form a circular patch antenna array of 
$\unit[8]{}$ elements with a distance between the adjacent antennas of 
$d$ = $\lambda/2$ (cf. Fig. \ref{fig:UCA}) in order to obtain the desired 
OAM wave having a radiation pattern of doughnut type 
with increased gain and lower side and back lobes. 
The slight asymmetry of the radiation pattern is due to 
the single-sided, hence asymmetric feeding of each patch 
antenna element. This asymmetry becomes more noticeable for 
larger radii. The reflection coefficient $S_{11}$ of the 
antennas are between $\unit[-17]{dB}$ and $\unit[-20]{dB}$ at the operating frequency 
of $\unit[10]{GHz}$. The gain of the planar circular patch antenna 
array amounts to about $\unit[9.5]{dBi}$ for the OAM mode $m=-1$. The footprint 
of the underlying PCB board is $\unit[100]{mm} \times \unit[100]{mm}$.
The UCA is designed with 
the full-wave simulator FEKO that applies the Method of Moments 
(MoM) providing a high simulation efficiency for this setup. 
The operating frequency is set to $\unit[10]{GHz}$. These antennas are 
atched on a Rogers RO4003C substrate with a  height of 
$\unit[1.524]{mm}$ and a relative permittivity of $\unit[3.55]{}$. 
The phase shift between each pair of adjacent 
antennas is defined as port feeding phase in the full-wave simulator FEKO to obtain 
the OAM mode with mode order $m=-1$. This phase shift is defined by the 
following relation

\begin{equation}
\varphi_1 =\frac{2\pi m}{N}, 
\end{equation}
where $N$ is the number of single antennas and 
$m$ is the mode order of the vortex waves.

\section{design of conventional and tailored lens}
\begin{figure}[H]
    \centering
    \includegraphics[width=80mm]{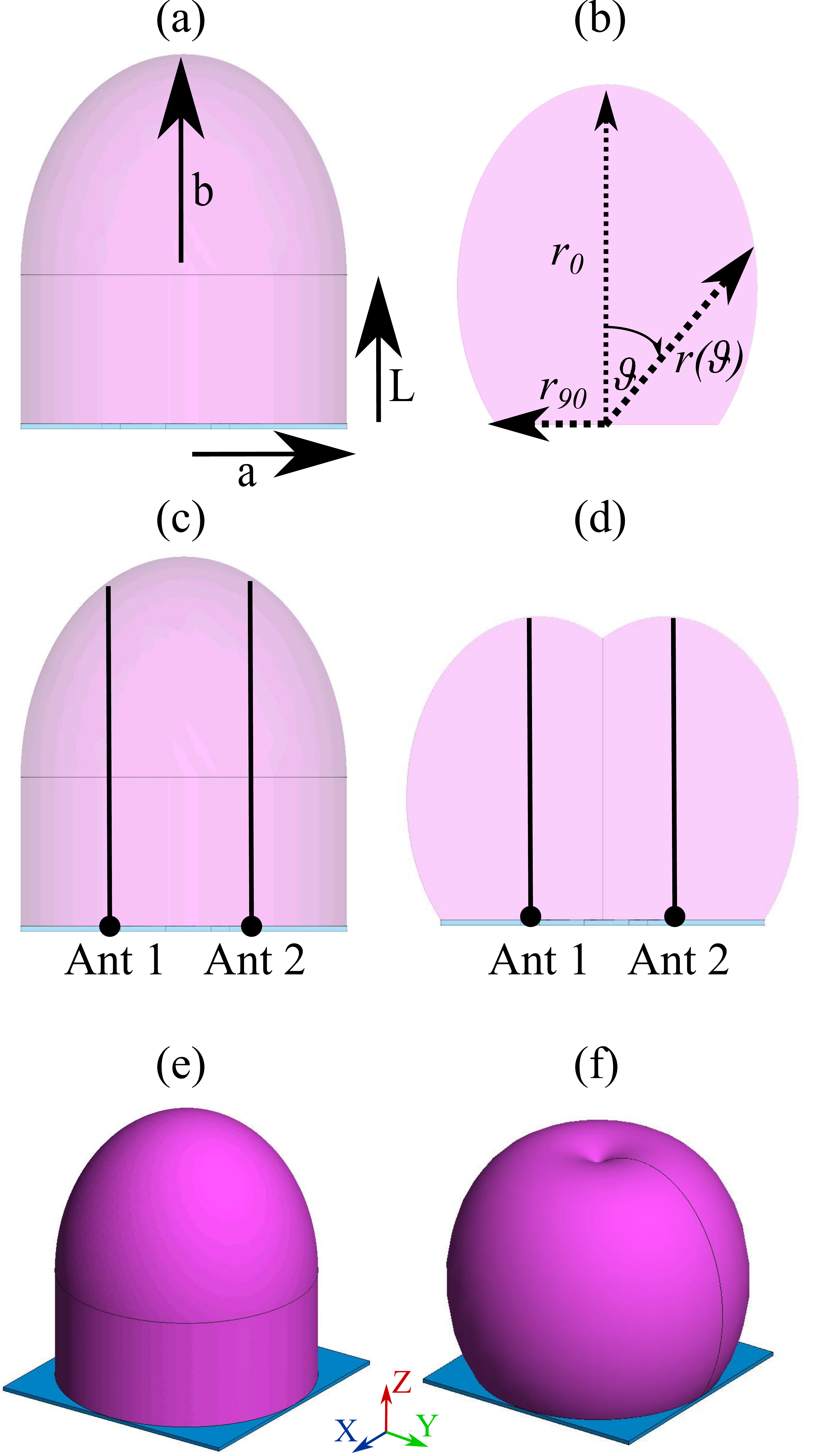}
    \caption{Conventional lens (a,c,e), and tailored lens (b,d,f).}
    \label{fig:Lens_Modell}
\end{figure}

In this Section, the conventional and the tailored lens are designed and compared to each other.
The material of the two lenses is polypropylene with a relative permittivity of $\unit[2.2]{}$.
Therefore, the antennas has to be redesigned accordingly due to the change
of the effective permittivity. 
The new length and the width of the patch antenna element are thus $\unit[7.25]{mm}$ and $\unit[10.2]{mm}$, respectively. 
The reflection coefficient $S_{11}$ amounts to $\unit[-37.6]{dB}$ for one patch antenna element at $\unit[10]{GHz}$. 
The conventional lens consists of two parts, namely the cylindrical dielectric part and the ellipsoidal part. 
The cylindrical dielectric part shifts the focal point into the center of the UCA
and the ellipsoidal part
converts the spherical waves into narrower plane waves.
Therefore, the gain of the UCA is expected to be increased and the divergence will be correspondingly reduced.
The design of the conventional lens uses the following equations \cite{hemishpericallens} 
according to Fig. \ref{fig:Lens_Modell} (a,c,e)

\begin{equation}
    b =\dfrac{a}{\sqrt{1-\dfrac{1}{\Re(\varepsilon_r)}}},
\end{equation}
    
    \begin{equation}
    L =\dfrac{b}{\sqrt{\Re(\varepsilon_r)}},
\end{equation}
    
    \begin{equation}
    D[dB] =20\log_{10}\left(\frac{2 \pi a}{\lambda}\right),
    \label{eq:Gain_conven}
\end{equation}

where $a$ and $b$ are the semi-minor and semi-major axis of the ellipsoidal part along the $z$-axis, 
where $a$ depends on the target directivity in [$\unit[]{dB}$].
$L$ is the length of the extension part, and $\lambda$ is the operating wavelength, which is $\unit[30]{mm}$ at $\unit[10]{GHz}$.
The equation (\ref{eq:Gain_conven}) seems to assume an aperture efficiency of $\unit[100]{}$\%, which
is most propably not true for OAM-antennas.
$\unit[50]{mm}$ is chosen for $a$ to make the assembling of the lens with the PCB board easier.
Hence, $L$ and $b$ are $\unit[45.6]{mm}$ and $\unit[67.7]{mm}$, respectively.
On the other hand, the principle of Fermat allows to design a lens for a single patch antenna
\cite{buchlens}

\begin{equation}
    r(\vartheta) =\left(\frac{r_0 (n_1-n_0)}{n_1-n_0 \cos(\vartheta)}\right),
    \label{eq:Fermat_lens}
\end{equation}

where $n_1$ and $n_0$ are the refractive indices 
of the lens and the air with values of $\unit[1.483]{}$ and $\unit[1]{}$, respectively.
$r$($\vartheta$) is the radius of the lens, which depends on the polar angle $\vartheta$,
and $r_0$ is the radius at $\vartheta$ = 0$^\circ$. 
The shape function (\ref{eq:Fermat_lens}) is used to design the tailored lens by sweeping 
this function which has been shifted to align with the pacth antenna's center arround
the $z$-axis, as shown in Fig. \ref{fig:Lens_Modell} (b,d,f).
Similar to the conventional lens, the entire PCB board is covered to simplify 
the assembling of the lens with the UCA.

\begin{figure}[H]
    \centering
    \includegraphics[width=80mm]{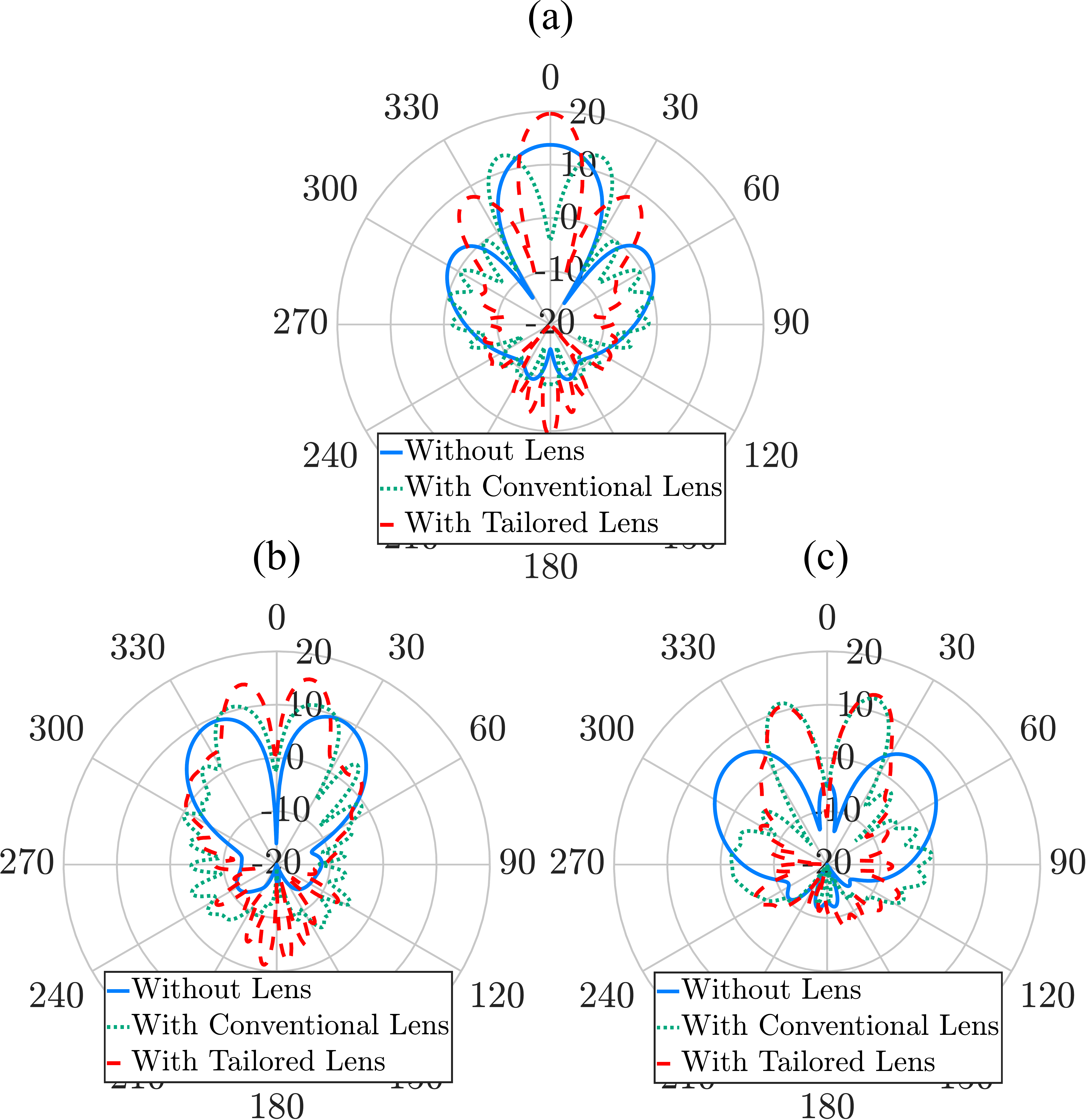}
    \caption{Radiation pattern of UCA at $\varphi  = 0^\circ$ for the case of without lens, with
    conventional lens and with tailored lens for the mode order $0$ (a), $-1$ (b), and $-2$ (c).}
    \label{fig:Alle_Moden_Lens}
\end{figure}

\begin{figure}[H]
    \centering
    \includegraphics[width=68mm]{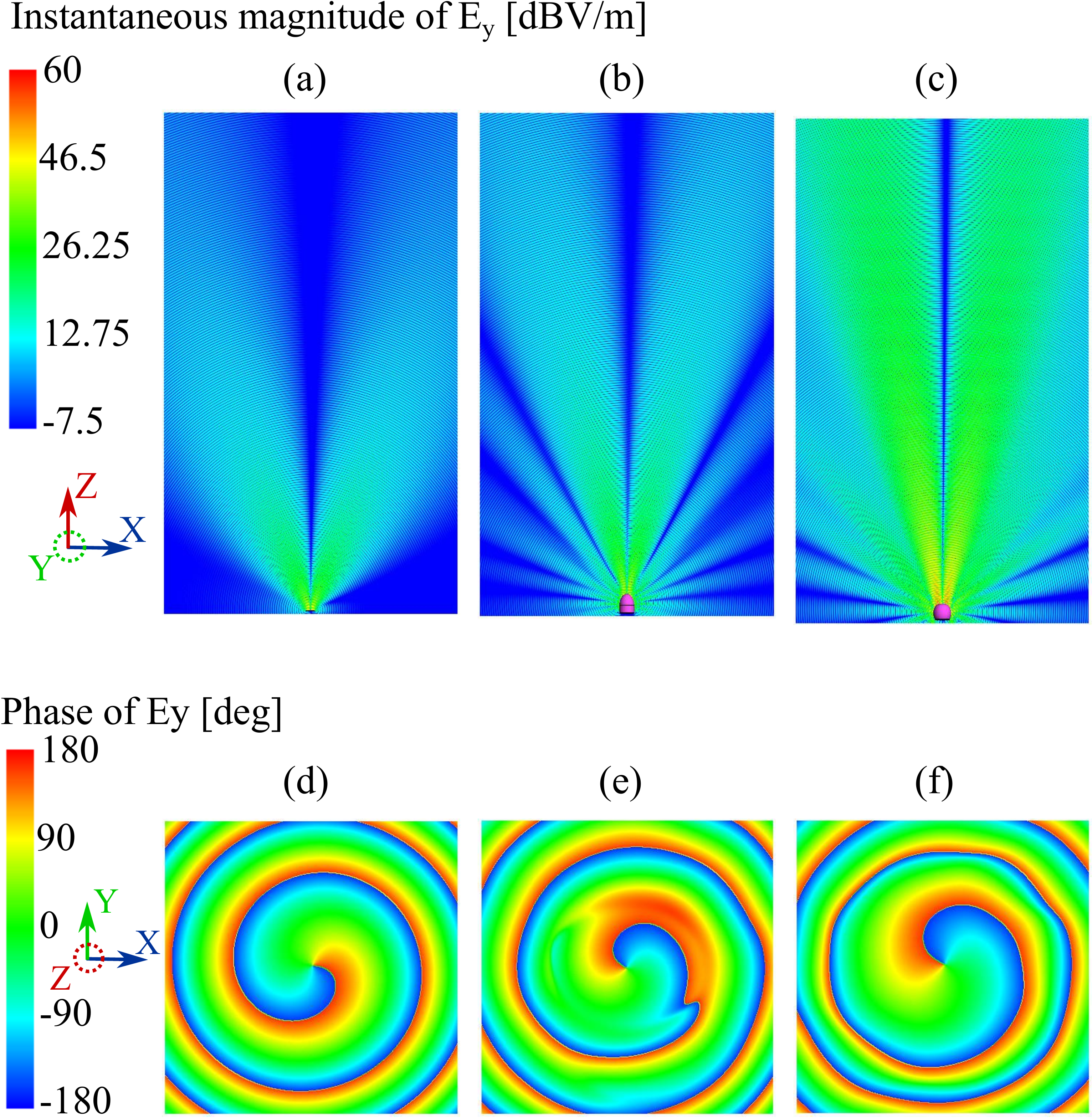}
    \caption{The instantaneous electric field amplitude and the phase distribution for the mode order $-1$ of the case without lens (a,d),
    with conventional lens (b,e), and with tailored lens (c,f), respectively.}
    \label{fig:Instantaneous_Electirc_Field_Lens}
\end{figure}

Thus, the radius at the polar angle $\vartheta$ $= 0$$^\circ$ and $\vartheta$ $= 90$$^\circ$ 
are about $\unit[93]{mm}$ and $\unit[30.4]{mm}$,
respectively. The larger $r_0$ is, the greater the gain of the radiation pattern and the divergence will be thus correspondingly reduced. 
The gain of the UCA for the mode order $-1$ is increased with the conventional lens from 
about $\unit[9.5]{dBi}$ to $\unit[11.3]{dBi}$ as depicted in Fig. \ref{fig:Alle_Moden_Lens} (b).
For an $a$ of $\unit[50]{mm}$ the gain shall be $\unit[20.4]{dBi}$ with a single antenna 
element. In addition to the increased gain, one can notice that the number of the side lobes is increased,
which leads automatically to reduction of the gain and the beam divergence.
The tailored lens (cf. Fig. \ref{fig:Alle_Moden_Lens} (b)) shows a significant decrease 
of the beam divergence with a maximum gain of $\unit[15.3]{dBi}$. 
Moreover, the number of the side lobes is less than in the case of the conventional lens.
One can notice that the aperture of the two lenses are not equal.
The effective aperture $A_e$, and the physical aperture $A_{phys}$ are defined with

\begin{equation}
    A_{e} =\frac{\lambda^{2}G}{4\pi},
    \label{eq:A_e}
\end{equation}

\begin{equation}
    A_{phys} =\pi r^2,
    \label{eq:A_phy}
\end{equation}

\begin{equation}
    e_{a} =\frac{A_{e}}{A_{phys}},
    \label{eq:A_phy_2}
\end{equation}
where $\lambda$ is the wavelength in the free space, $G$ is the gain of the antenna,
and $r$ is the largest radius of the lens.
The ratio between the effective aperture and the physical aperture is called 
the aperture efficiency $e_a$, which is a parameter between $0$ and $1$,
that measures how good is the antenna by receiving the radio wave power, which enters in the physical aperture.
The conventional lens and the tailored lens have an aperture efficiency $e_a$ of $0.12$ and $0.21$, respectively.
This is an advantage of the tailored lens compared to the conventional lens.
In Fig. \ref{fig:Alle_Moden_Lens} (a), the mode order $0$ is depicted.
One can notice that the conventional lens is deforming the mode $0$ contrary to the tailored lens,
which is enhancing the gain from $\unit[13.7]{dBi}$ to $\unit[19.5]{dBi}$.
This is due to the placement of the antennas, which are not in the focal point in the case of the conventional lens.
This is a very good advantage of the tailored lens compared to the conventional lens, especially
when the lens is used in the OAM target localization, where several mode orders are needed to 
find the direction of the target.
In Fig. \ref{fig:Alle_Moden_Lens} (c), in the case of mode $\unit[-2]{}$ (cf. Fig. \ref{fig:Alle_Moden_Lens} (b)) 
the two lenses have carried out a similar gain enhancement, but with an advantage of less side lobes in the case of the tailored lens.
Please note that the maximum reached gain with the second mode order can be achieved with a distance 
$d$ of about $\lambda$ between the adjucent antenna element (cf. Fig. \ref{fig:UCA} (a)).
Therefore, this leads to increase the size of the lens and to use more material, which is not the case 
of the tailored lens, which can save material and weight especially in the middle.
This is an additionally advantage of the tailored lens compared to the conventional lens.
Fig. \ref{fig:Instantaneous_Electirc_Field_Lens} shows the instantaneous electric field and 
the phase distribution of the OAM mode order $\unit[-1]{}$ of the case 
without lens (a,d), with conventional lens (b,e), and with tailored lens (c,f).


\section{Lens fabrication and measurements}

\begin{figure}[H]
    \centering
    \includegraphics[width=90mm]{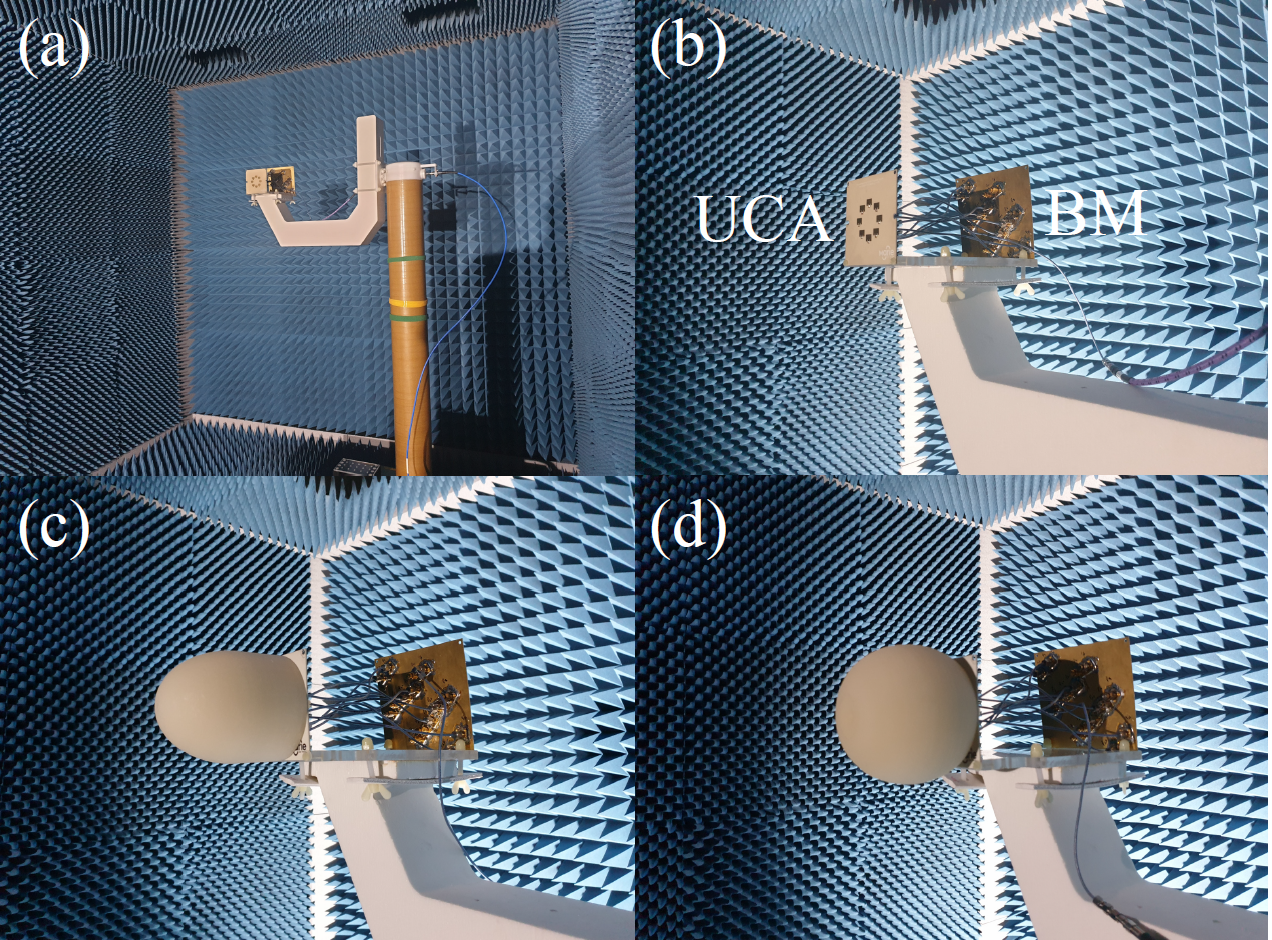}
    \caption{The rotary table in the anechoic chamber (a), the manufactured UCA with BM (b), the conventional lens on the UCA (c),
    and the tailored lens on the UCA (d).}
    \label{fig:Lens_Real}
\end{figure}

The designed conventional lens and tailored lens in the previous section
have been manufactured by an external company.
The measurement is performed in an anechoic chamber in order to avoid 
unwanted reflections and distortions.
A vector network analayzer ZVA $40$ from Rohde \& Schwarz is used at $\unit[10]{GHz}$.
The VNA is in a control room and connected with the antennas
by two coaxial cables with a length of $\unit[5]{m}$.
A horn antenna is used as a transmitter and the lens with UCA and the Butler matrix (BM),
which is providing different OAM mode orders, are used as a receiver.
The distance between the transmitter and the receiver is about $\unit[5]{m}$. 
The lens with UCA and BM are mounted on a rotary table,
which is rotating in the azimuth angle $\varphi$ from $0^\circ$ to $180^\circ$ and in the 
elevation angle $\vartheta$ from $-45^\circ$ till $45^\circ$.
The provided mode order $\unit[1]{}$ is provided by the $8$$\times$$8$ BM, which is connected to the UCA 
by eight coaxial cables of equal length of $\unit[200]{mm}$ (cf. Fig. \ref{fig:Lens_Real}).
The two lenses are assembled separately on the UCA.

\begin{figure}[H]
    \centering
    \includegraphics[width=90mm]{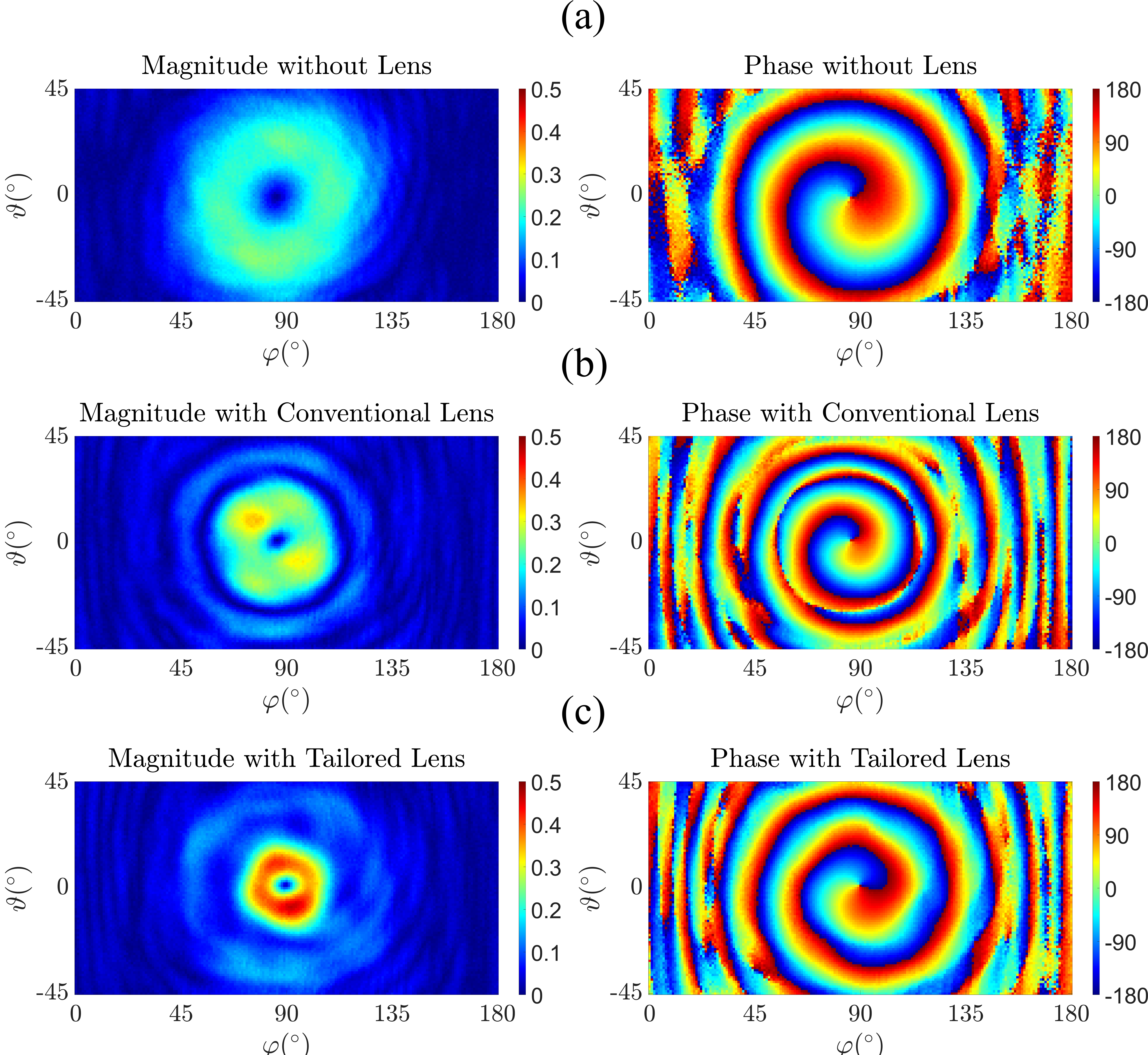}
    \caption{The amplitude and the phase distribution for the mode order $1$ of antennas without lens (a),
    with conventional lens (b), and with tailored lens (c).}
    \label{fig:Lens_Messung_3d}
\end{figure}

\begin{figure}[H]
    \centering
    \includegraphics[width=90mm]{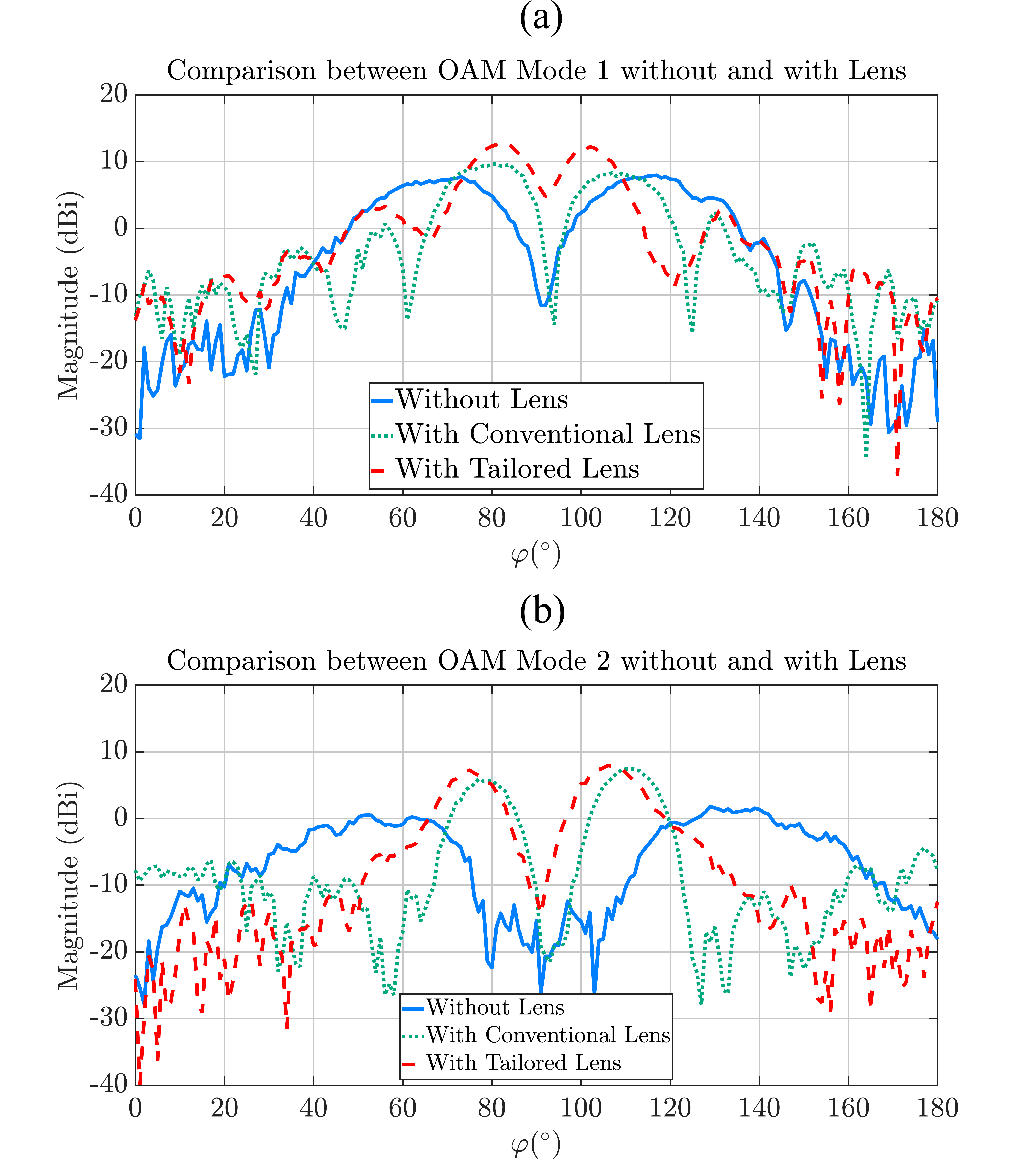}
    \caption{Comparison between the antennas at $\vartheta=0^\circ$ without lens, with conventional lens, 
    and with tailored lens for the mode order $1$ (a), and $2$ (b).}
    \label{fig:Lens_Messung_2d}
\end{figure}

In Fig. \ref{fig:Lens_Messung_3d}, the gain of the case 
without lens (a), with conventional lens (b), and with tailored lens (c) are depicted, respectively.
It is easy to observe that the tailored lens has a much better performance 
compared to the two other cases.
The beam divergence is reduced,
furthermore, the gain is increased from $\unit[8]{dBi}$ 
to $\unit[9.7]{dBi}$ and from $\unit[8]{dBi}$ to $\unit[12.8]{dBi}$ 
for the case of the conventional and the tailored lens, respectively.
Fig. \ref{fig:Lens_Messung_3d} shows also the phase distribution of the three cases.
The three cases are showing a phase distribution of a one helix,
which is a sign of the first mode order.
The number of the helices indicates the mode order and the direction of the rotation
indicates the sign of the mode order.
The sign is positive when the helix is rotating clockwise in the propagation direction
and it is negative when the helix is rotating counterclockwise.
Fig. \ref{fig:Lens_Messung_2d} shows the magnitude of the three cases for mode $\unit[1]{}$ and $\unit[2]{}$, respectively.
One can remark that the gain in the center of of the doughnut is not $0$ (linear), like the ideal case.
This is due to many reasons, such as the not exact alignment between the transmitter and the receiver,
and due to the reflections in the BM and the cables, which are not also ideal.
One can also see that the higher the mode order is, the larger the vortex beam is.

\section{design of conventional and tailored reflector with impressed field source}

Likewise the lens, Fermat's principle allows for a conventional 
reflector for a point source \cite{buchlens} to be designed, 
where $r$($\vartheta$) is the radius of the reflector depending
on the polar angle $\vartheta$, $r_0$ is the radius at
$\vartheta$ = 0$^\circ$, $n_1$ is the refractive index 
of the air with a value of 1 

\begin{equation}
r(\vartheta) =\left(\frac{2 r_0 }{n_1(1+\cos(\vartheta))}\right).
\label{eq:Fermat_Reflector}
\end{equation}

\begin{figure}[H]
    \centering
    \includegraphics[width=80mm]{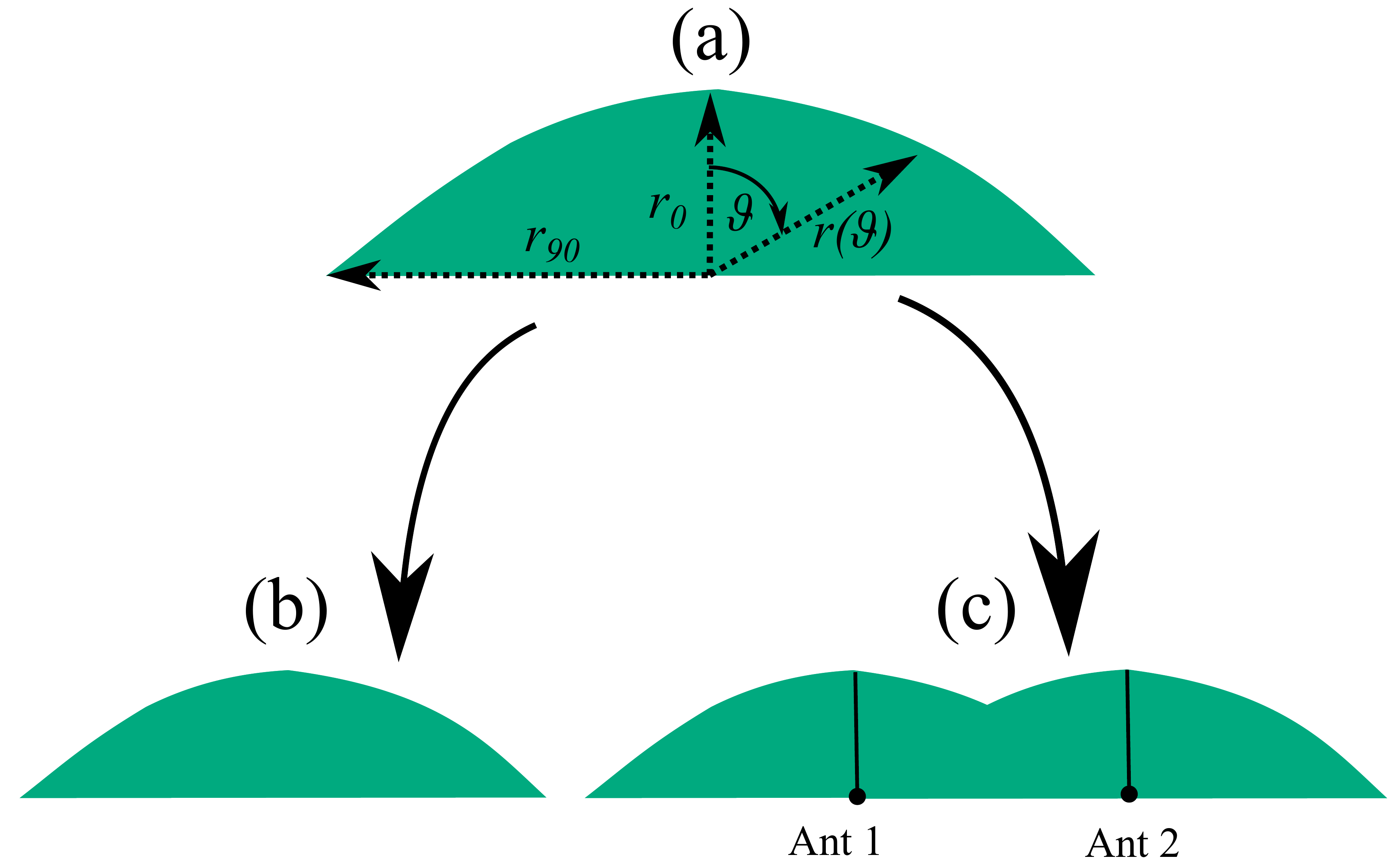}
    \caption{Reflector for a point source (a), extension to conventional reflector (b),
    and extension to tailored reflector (c).}
    \label{fig:Reflector_modell}
\end{figure}

\begin{figure}[H]
    \centering
    \includegraphics[width=80mm]{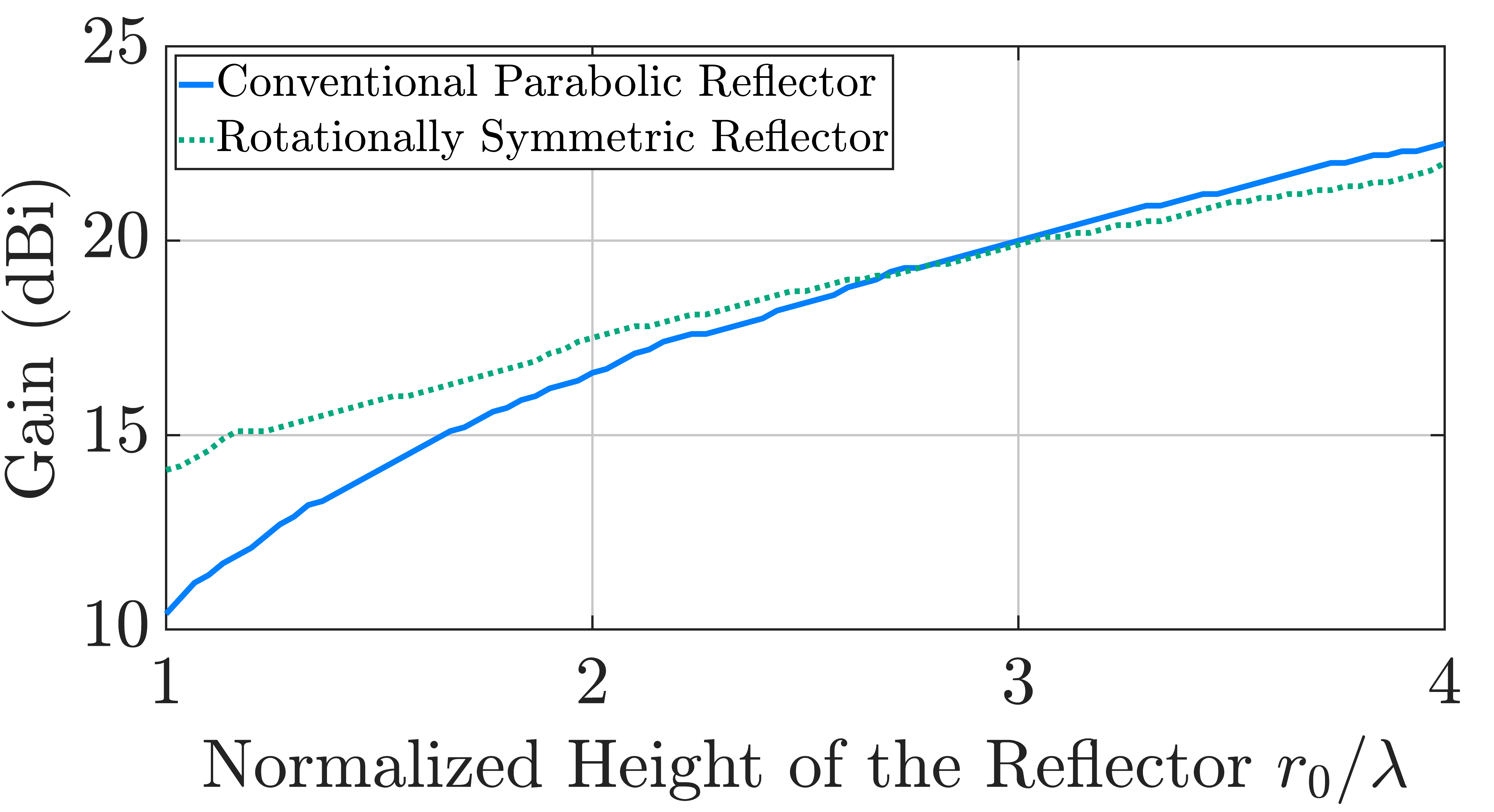}
    \caption{Gain for the mode order $-1$ depending on the height $r_0$ of the reflector.}
    \label{fig:Gain_r}
\end{figure}

The tailored reflector is designed by sweeping 
the shape function which also has been shifted to align with the pacth antenna's center arround
the $z$-axis (\ref{eq:Fermat_Reflector}),
as shown in Fig. \ref{fig:Reflector_modell} (a,c).
The two reflectors depend on two parameters, the radius 
$r$ and the angle $\vartheta$. 
The radius depends on how large the reflector is and on the expected gain. 
A large reflector leads to higher gain.
Consequently, the divergence of the vortex waves will be reduced
due to an increased focusing of the radiation pattern.
The angle $\vartheta$ is adjuted from
$-90$$^\circ$ till $90$$^\circ$. The tailored 
reflector will be compared to UCA without reflector 
and to UCA with conventional reflector. In Fig. \ref{fig:Gain_r} 
the gain of the two reflector depending on the height $r_0$ is depicted, where
we observe that the performance of the tailored reflector is better 
than the conventional reflector, especially till the 
height of $1.5$$\lambda$. Beyond $1.5$$\lambda$ there is no huge gain difference 
between the conventional reflector and the tailored reflector. 
However, the tailored reflector continues to show a better reduction 
in the divergence because the maximum gain still close to the 
center of  the OAM  waves (cf. table \ref{tab:table1}).
These results are derived without the influence of the 
circular antenna array, which may cause some reflections
and diffractions of the reflected waves.
Therefore, it is easier to 
compare the two reflectors.

\begin{table}[htpb]
    \caption{Opening angle where the maximum Gain of each reflectors
    ($\lambda_0$$=30 mm$ at $\unit[10]{GHz}$) for the mode order $-1$.}
    \label{tab:table1}
    \centering
    \begin{tabular}{ c | c | c }
        $r_0$ & Opening angle {(}$^\circ${)} & Opening angle {(}$^\circ${)} \\
            (mm) & (Conventional reflector) & (Tailored reflector)  \\
    \hline
    $30$                       & $155$        &  $168$     \\   
    $32$                       & $156$        &  $169$     \\  
    $34$                       & $157$        &  $170$     \\  
    $42$                       & $161$        &  $171$     \\  
    $50$                       & $164$        &  $172$     \\ 
    $59$                       & $166$        &  $173$     \\ 
    $72$                       & $169$        &  $174$     \\ 
    $100$                      & $172$        &  $175$     \\
    $113$                      & $173$        &  $176$     \\ 
\end{tabular}
\end{table}

\begin{figure}[H]
    \centering
    \includegraphics[width=90mm]{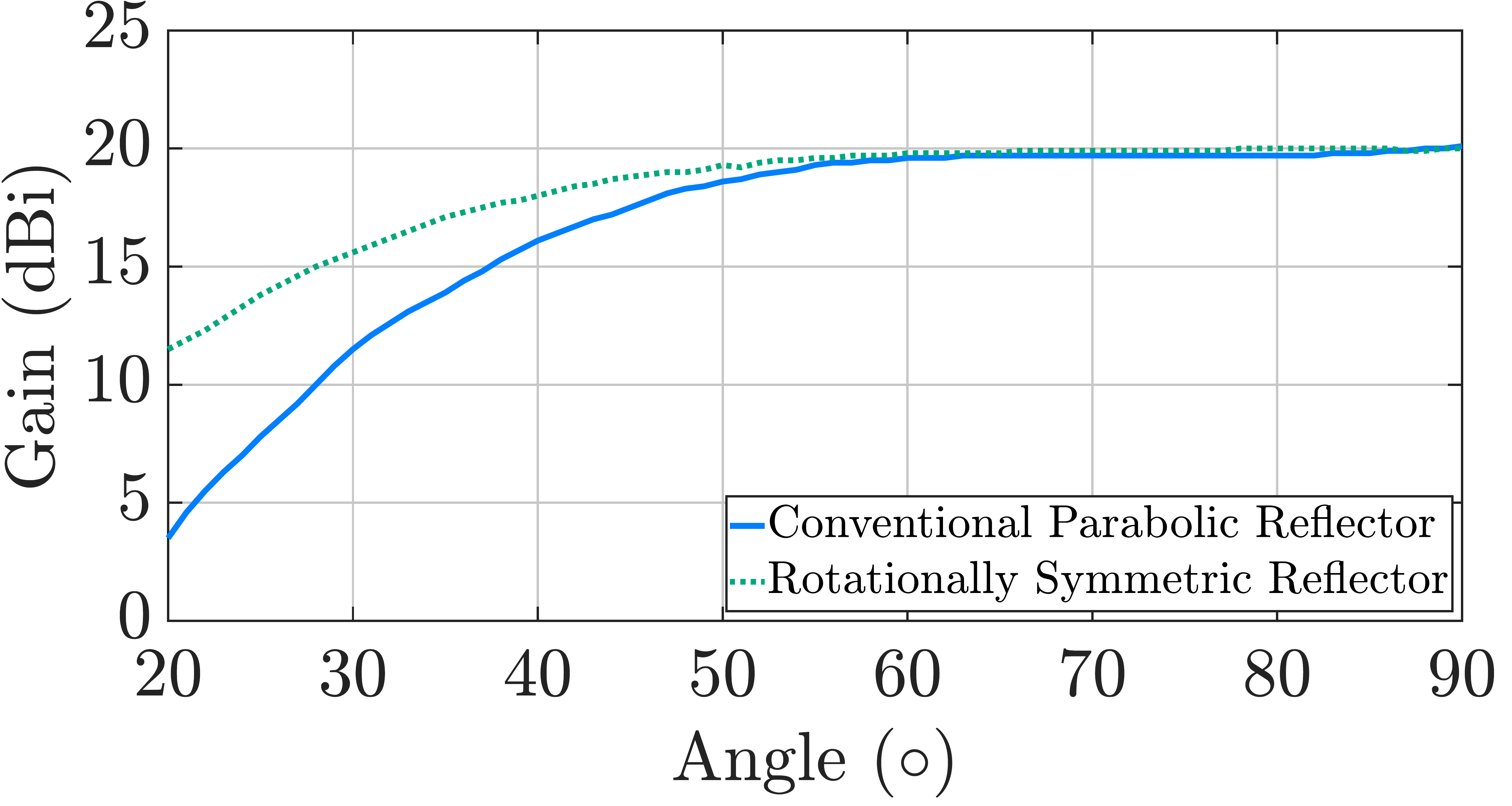}
    \caption{Gain for the mode order $-1$ depending on the angle $\vartheta$ with a height $r_0$ of $\unit[90]{mm}$ of impressed 
    field source with conventional reflector and  with tailored reflector.}
    \label{fig:Gain_winkel}
\end{figure}

\begin{figure}[H]
    \centering
    \includegraphics[width=70mm]{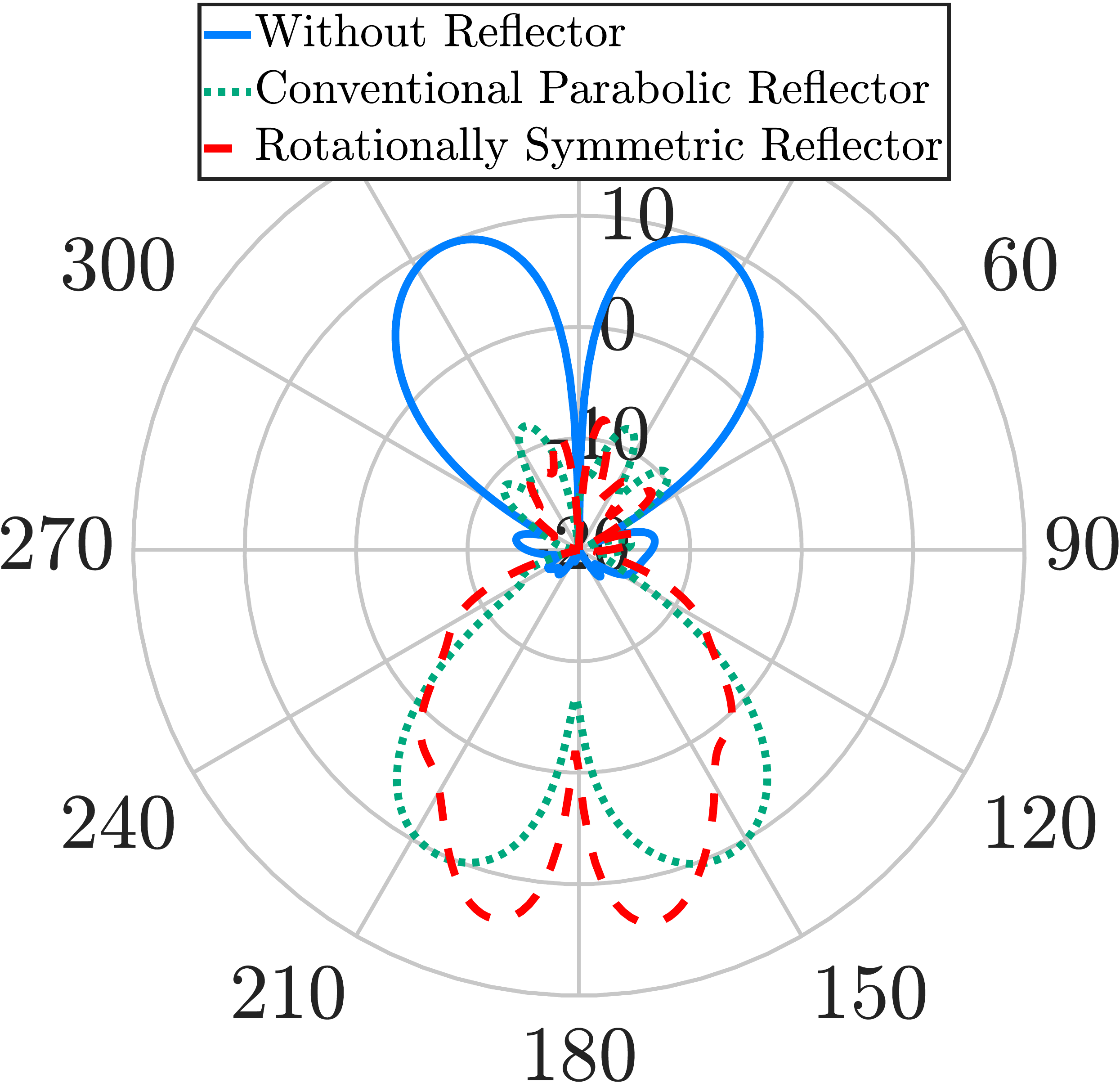}
    \caption{Radiation pattern of impressed field source for the mode order $-1$ with $r_0$ of $\unit[40]{mm}$ and with an angle $\vartheta$
    from $-90$$^\circ$ till $90$$^\circ$ without reflector, 
    with conventional reflector and with tailored reflector.}
    \label{fig:RP_im}
\end{figure}

\begin{figure}[H]
    \centering
    \includegraphics[width=65mm]{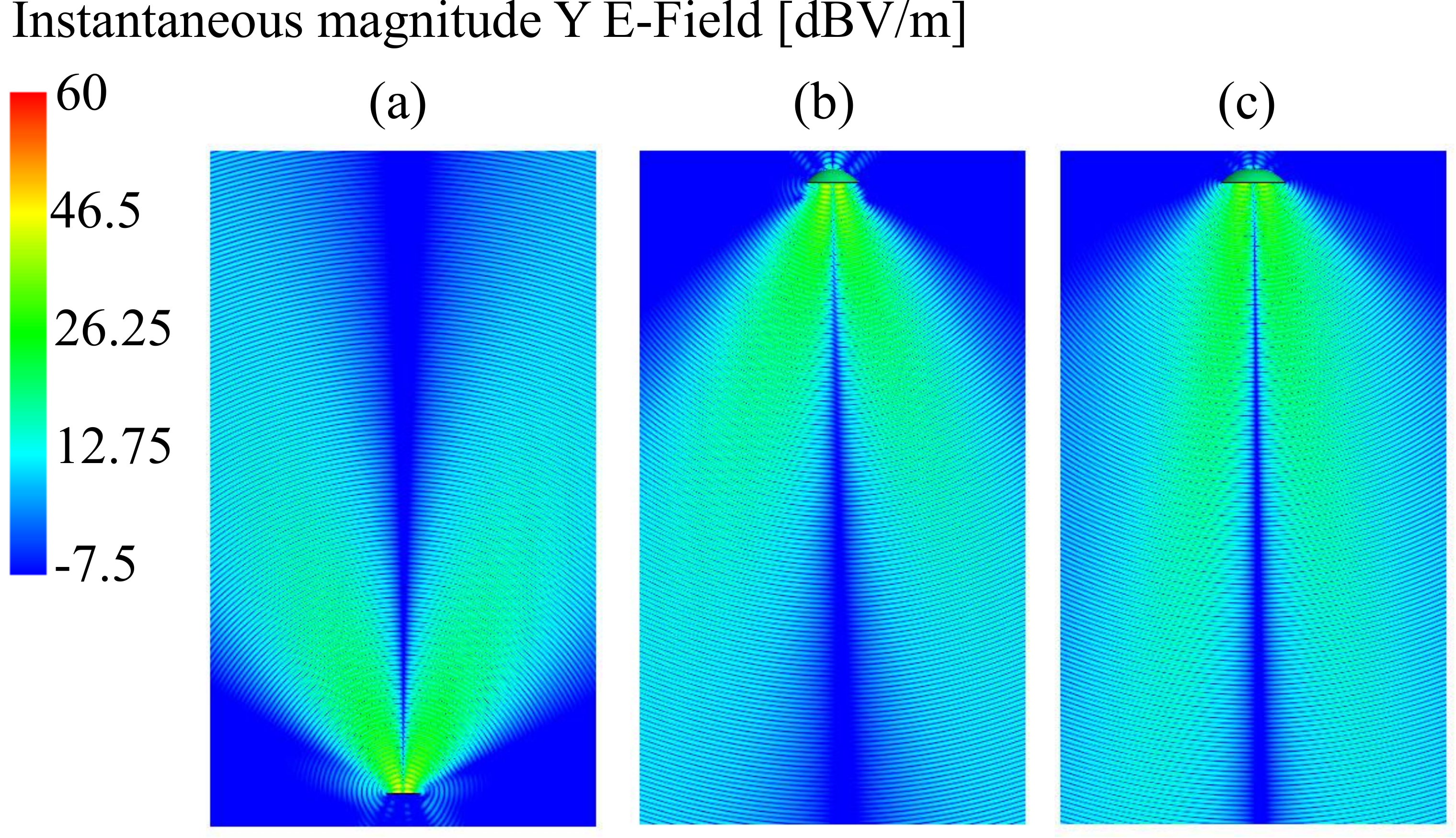}
    \caption{The instantaneous electric field amplitude of impressed field source for the mode order $-1$ with $r_0$ of $\unit[40]{mm}$ and with an angle $\vartheta$
    from $-90$$^\circ$ till $90$$^\circ$ without reflector (a), 
    with conventional reflector (b) and with tailored reflector (c).}
    \label{fig:Inst_im}
\end{figure}

\begin{figure}[H]
    \centering
    \includegraphics[width=65mm]{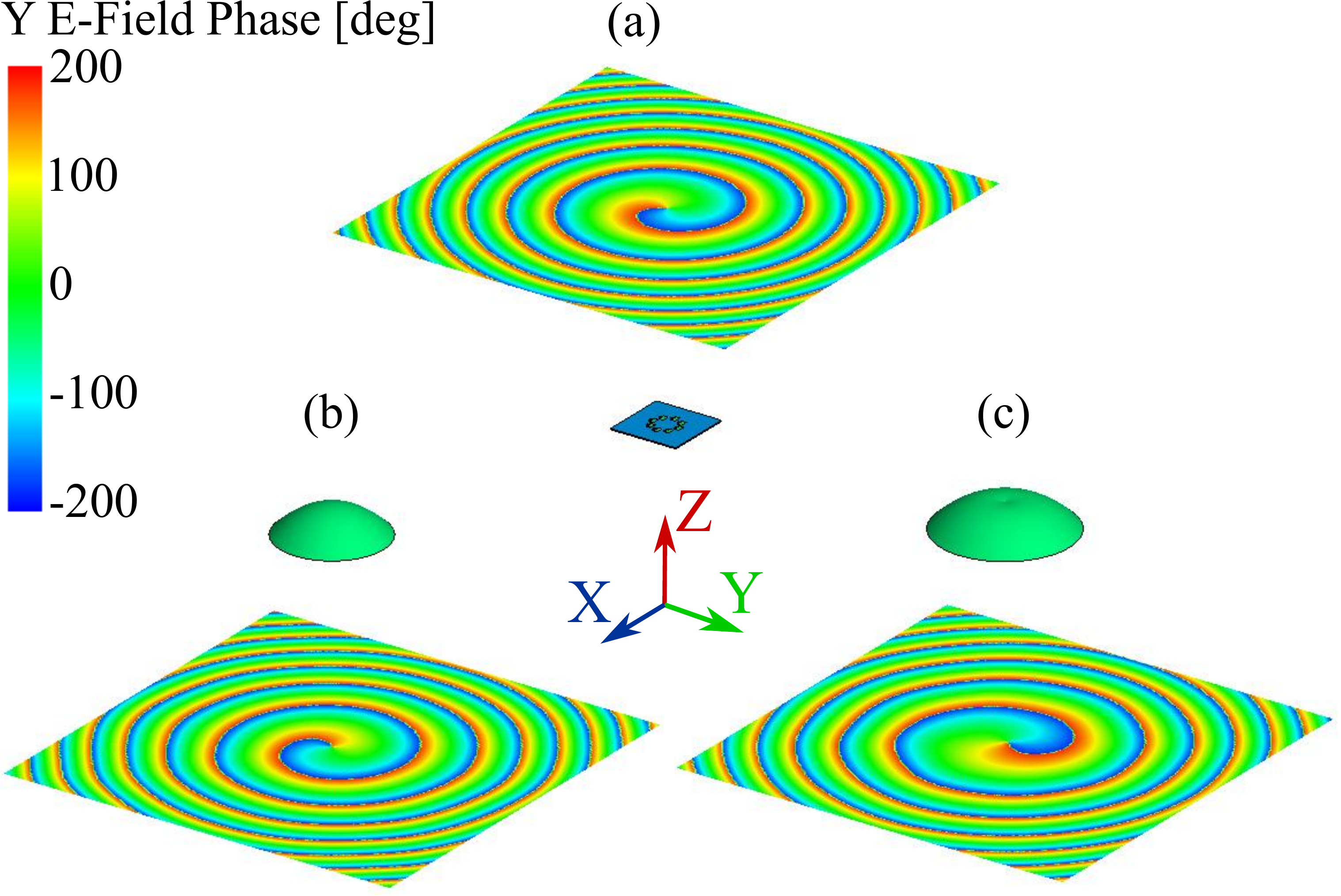}
    \caption{Phase distribution of $E_y$ of impressed field source for the mode order $-1$ ($x = -300$ till $\unit[300]{mm}$, $y = -300$ till 
    $\unit[300]{mm}$, $z = \unit[300]{mm}$ for (a) and $z = \unit[-300]{mm}$ for (b and c)) indicating 
    the phase distribution of impressed field source with $r_0$ of $\unit[40]{mm}$ and with an angle $\vartheta$
    from $-90$$^\circ$ till $90$$^\circ$ without reflector (a), 
    with conventional reflector (b) and with tailored reflector (c).}
    \label{fig:Phase_im}
\end{figure}

In Fig. \ref{fig:Gain_winkel} the two reflectors are 
compared for several angle $\vartheta$. As we can see 
the tailored reflector has also a better performance till 
the angle $\vartheta$ of $38$$^\circ$. 
In Fig. \ref{fig:RP_im} the simulated radiation pattern of the
impressed field source at $\varphi $ = 0$^\circ$ without reflector, with
conventional reflector 
and with tailored reflector are depicted.
The height $r_0$ is $\unit[40]{mm}$ and the angle $\vartheta$
is from $-90$$^\circ$ till $90$$^\circ$.
The two reflectors are reducing the divergence from about $\unit[9.5]{dBi}$ at angle $336$$^\circ$
to $\unit[13.2]{dBi}$ (conventional) at angle $160$$^\circ$ and to $\unit[15.4]{dBi}$ (tailored) at angle $170$$^\circ$.
One can notice that the tailored reflector is causing
some enlargement.
This is mainly due to the cut of the reflector in the center, which
let some rays to propagate into the second part of the reflector, where
undesired reflections can occur.
This will not occur when the radius of the UCA is too big compared to the $r_0$ of the reflector.
In Figs. \ref{fig:Inst_im} and \ref{fig:Phase_im} the instantaneous electric field and
the phase distribution (helical) of the $3$ cases are presented. 
The vortex waves are still achieved after the reflection with the 
reflectors, but with the opposite
mode, namely $+1$ instead of $-1$. The height $r_0$ of the reflectors is
$\unit[40]{mm}$ and the angle $\vartheta$
is from $-90$$^\circ$ till $90$$^\circ$.

\section{reflector evaluation including real feeding antenna structure}

\begin{figure}[H]
    \centering
    \includegraphics[width=85mm]{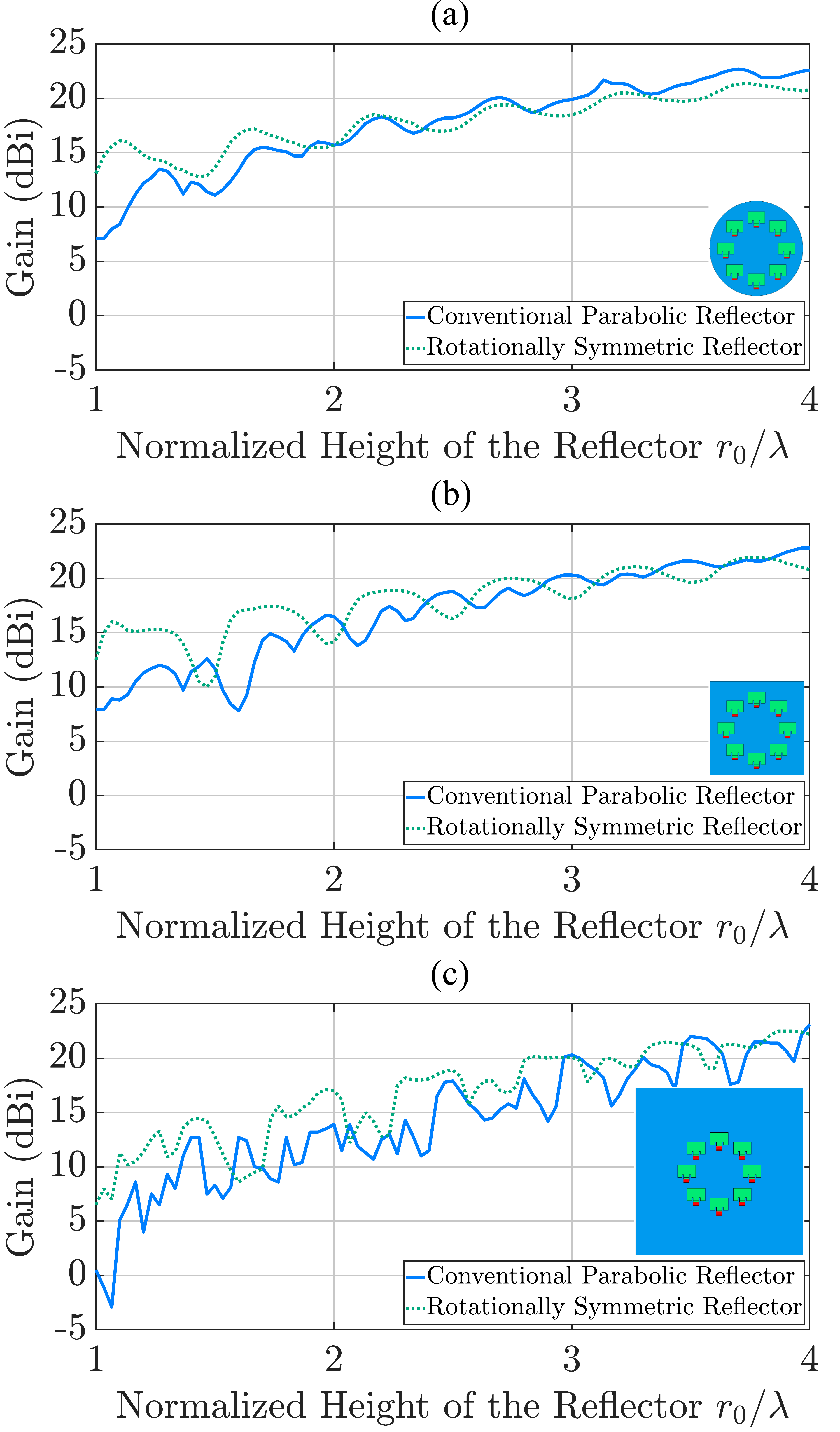}
    \caption{Gain for the mode order $-1$ depending on the height $r_0$ of the reflector for circular shaped 
    PCB with a diameter of $\unit[60]{mm}$ (a), for rectangular shaped PCB $\unit[60]{mm} \times \unit[60]{mm}$ (b) and for rectangular 
    shaped PCB with $\unit[100]{mm} \times \unit[100]{mm}$ (c).}
    \label{fig:Gain_r_real}
\end{figure}

\begin{figure}[H]
    \centering
    \includegraphics[width=60mm]{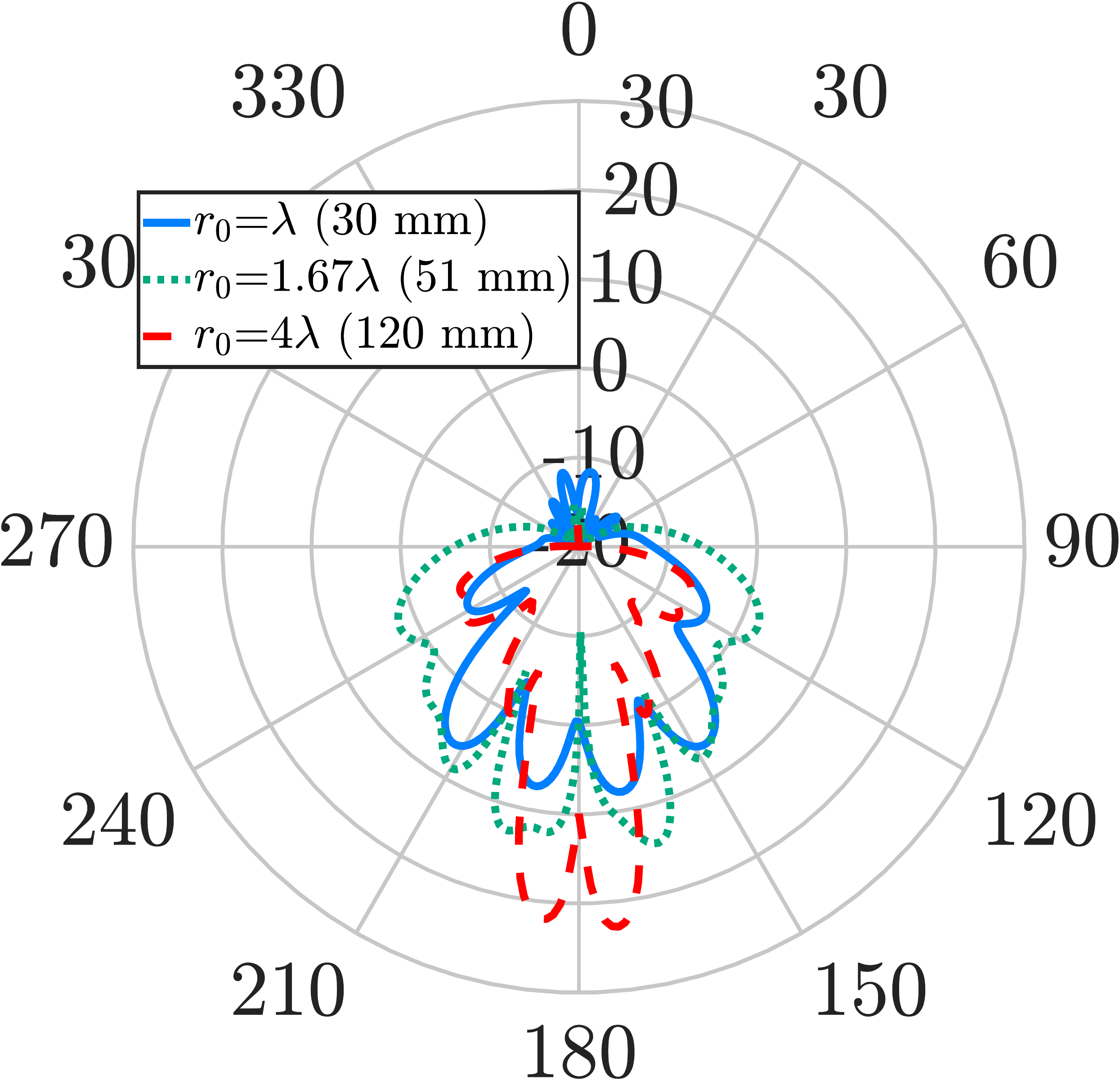}
    \caption{Radiation pattern with real UCA for the mode order $-1$ of conventional reflector of UCA 
    with a rectangular shaped PCB $\unit[60]{mm} \times \unit[60]{mm}$ for several 
    height $r_0$ of $30$, $51$, and $\unit[120]{mm}$.}
    \label{fig:RP_conventional}
\end{figure}

\begin{figure}[H]
    \centering
    \includegraphics[width=85mm]{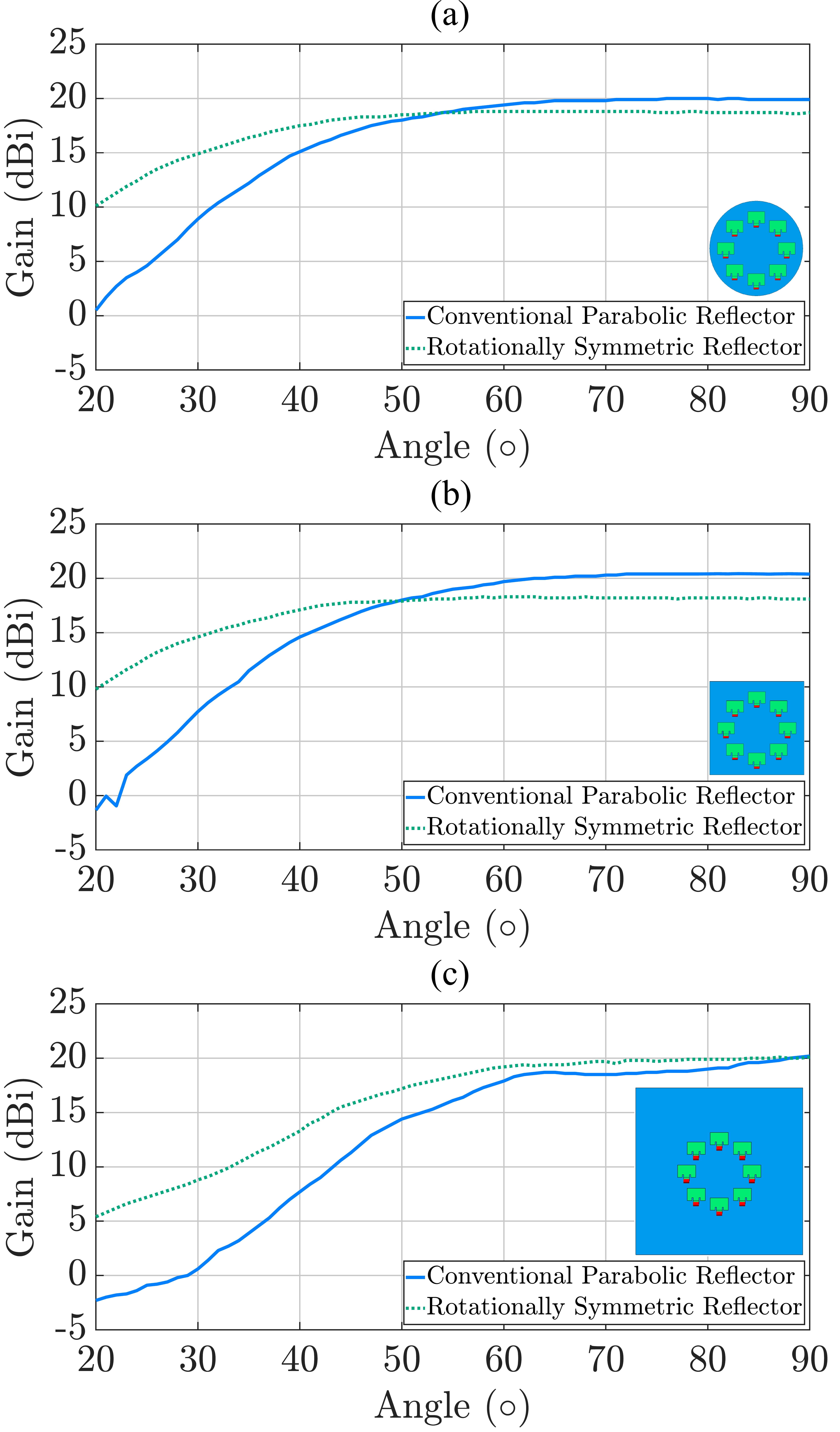}
    \caption{Gain for the mode order $-1$ depending on the angle $\vartheta$ of the reflector for the mode order $-1$
    with a height $r_0$ of $\unit[90]{mm}$
    for circular shaped PCB with a diameter of $\unit[60]{mm}$ (a), for rectangular shaped PCB with $\unit[60]{mm} \times \unit[60]{mm}$ 
    (b), and for rectangular shaped PCB with $\unit[100]{mm} \times \unit[100]{mm}$ (c).}
    \label{fig:Gain_win_real}
\end{figure}

\begin{figure}[H]
    \centering
    \includegraphics[width=59mm]{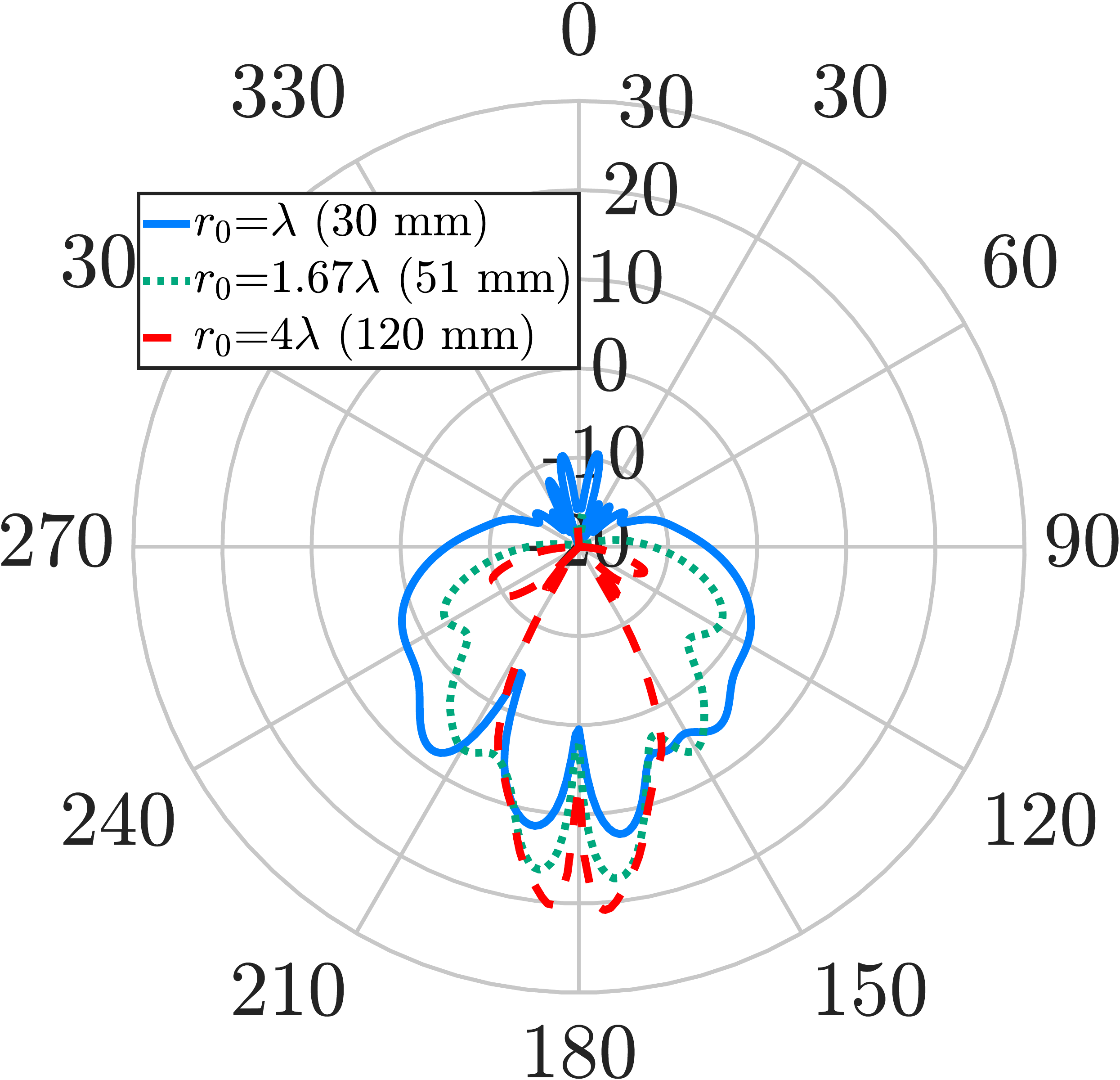}
    \caption{Radiation pattern with real UCA for the mode order $-1$ of tailored reflector of UCA with a rectangular shaped form $\unit[60]{mm} \times \unit[60]{mm}$ for several 
    height $r_0$ of $30$, $51$, and $\unit[120]{mm}$.}
    \label{fig:RP_tailored}
\end{figure}

\begin{figure}[H]
    \centering
    \includegraphics[width=60mm]{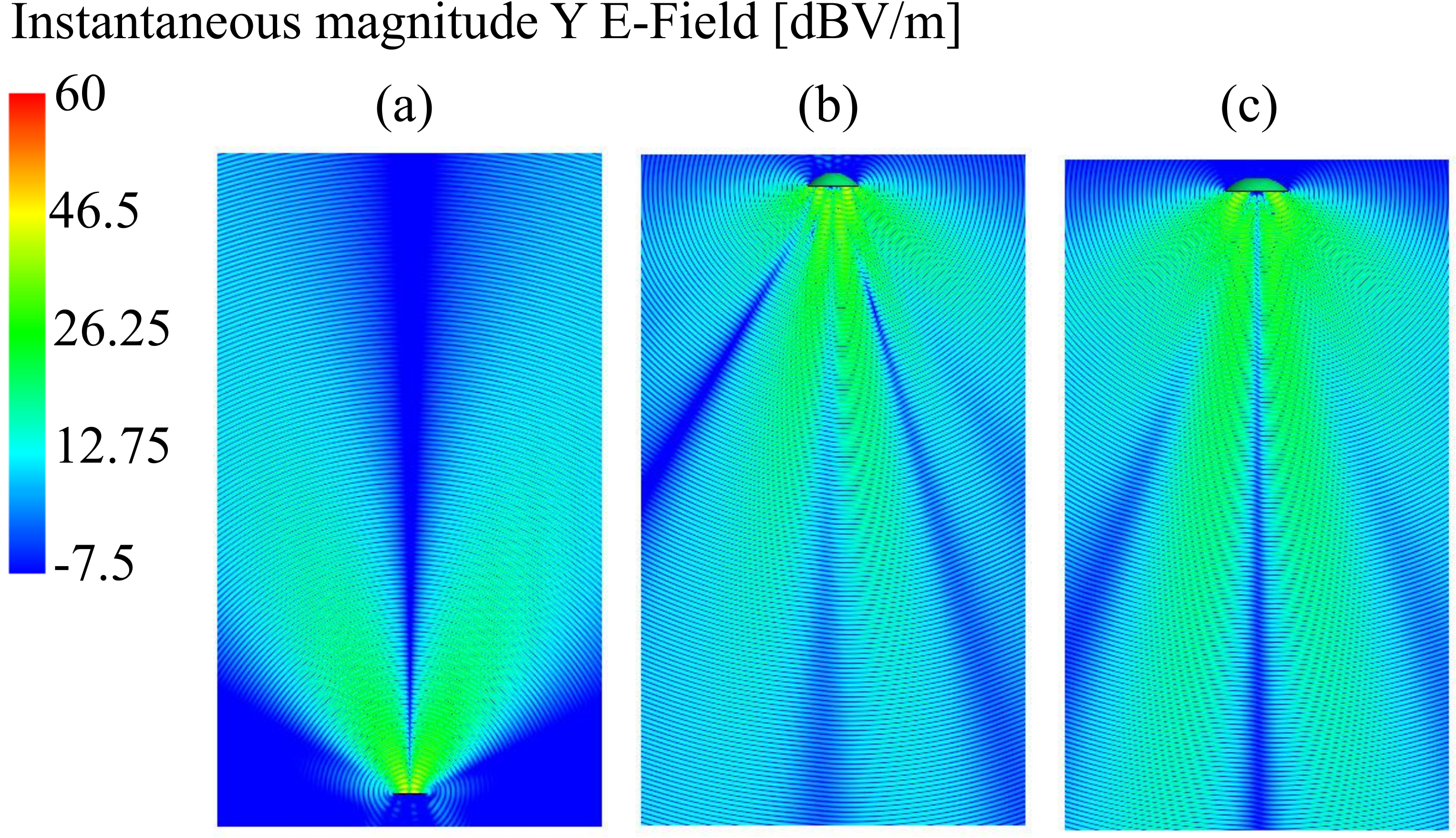}
    \caption{The instantaneous electric field amplitude with real UCA for the mode order $-1$ of the circular antenna 
    array with rectangular shaped PCB $\unit[60]{mm} \times \unit[60]{mm}$ without reflector (a), 
    with conventional reflector (b), and with tailored reflector (c).}
    \label{fig:ins_real}
\end{figure}

\begin{figure}[H]
    \centering
    \includegraphics[width=60mm]{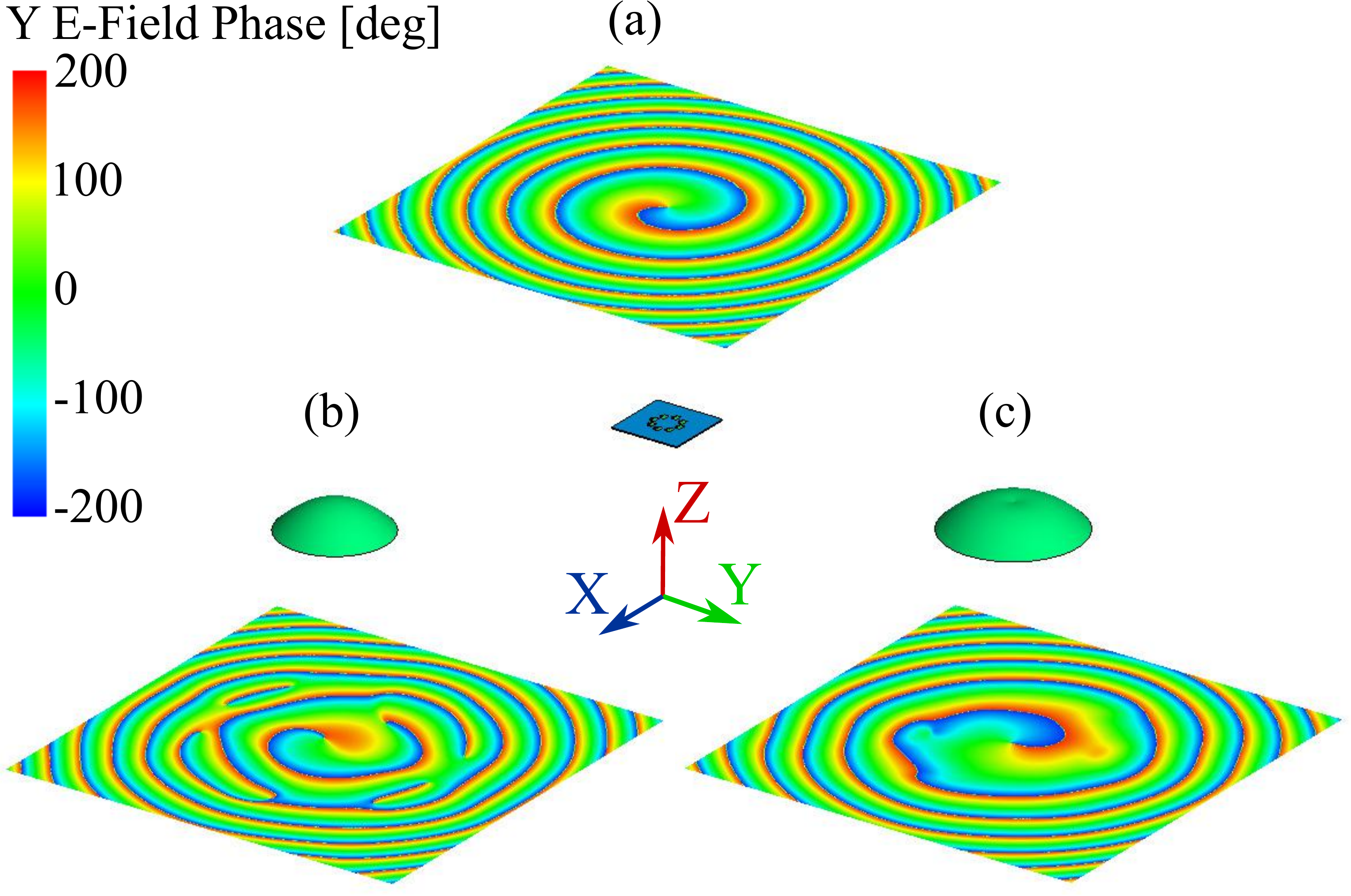}
    \caption{Phase distribution of $E_y$ with real UCA for the mode order $-1$ with rectangular shaped PCB $\unit[60]{mm} \times \unit[60]{mm}$
    ($x = -300$ till $\unit[300]{mm}$, $y = -300$ till $\unit[300]{mm}$, $z = \unit[300]{mm}$ for (a) and $z = \unit[-300]{mm}$ for (b and c)) indicating the 
    helical phase distribution of circular antenna array without and with reflector.}
    \label{fig:phase_real}
\end{figure}

\begin{figure}[H]
    \centering
    \includegraphics[width=60mm]{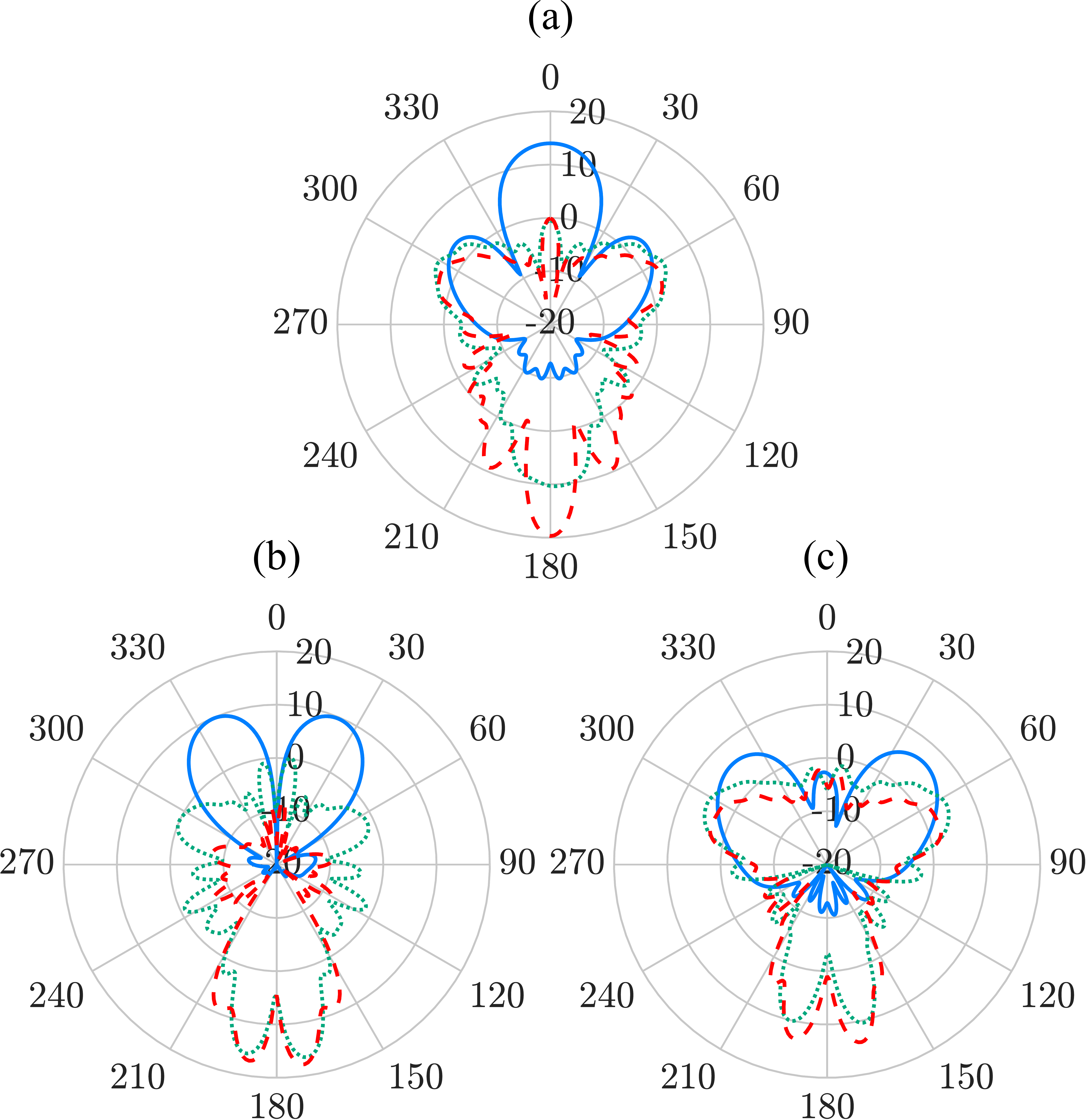}
    \caption{Radiation pattern for the three cases for the mode order $0$ (a), $-1$ (b), and $-2$ (c)
    with a height of $\unit[90]{mm}$ and an angle of $45$$^\circ$.}
    \label{fig:allemode_real}
\end{figure}

In the last section, the $3$ cases are presented with impressed field source, 
to simplify the interpretation of the behavior of the reflectors. 
In this section, the reflectors are simulated with real UCA.
Three different models are designed.
The first model is a circular shaped PCB with a diameter of $\unit[60]{mm}$.
The second and the third model have a rectangular shaped PCB with 
$\unit[60]{mm} \times \unit[60]{mm}$ and $\unit[100]{mm} \times \unit[100]{mm}$ size, respectively. 
Fig. \ref{fig:Gain_r_real} shows the gain depending on the height $r_0$.
One can recognize the influence of the 
reflector height. The oscillating of the gain is 
due to the reflection 
and the diffraction on the board of the transmitter.
Standing waves occur between the UCA and the reflector. 
The conventional reflector has more negative influence 
on the patch antenna than 
the tailored reflector, which is an advantage for the tailored 
reflector.
The circular shaped PCB has less influence on the vortex waves.
In Fig. \ref{fig:Gain_win_real} the gain depending on the angle $\vartheta$
with a height $r_0$ of $\unit[90]{mm}$ is presented.
In Figs. \ref{fig:RP_conventional} and \ref{fig:RP_tailored} the radiation pattern in $2$D 
at $\varphi $ = 0$^\circ$ of the conventional reflector and
the tailored reflector for the mode order
$-1$ for three different 
heights $r_0$ of $30$, $51$, and $\unit[120]{mm}$ are presented. 
The tailored reflector works well from the height $r_0$ of $\lambda$ 
contrary to the conventional reflector, which is showing a nice 
OAM beam from the height $r_0$ of 1.67$\lambda$.
In Figs. \ref{fig:ins_real} and \ref{fig:phase_real} the instantaneous electric field and the phase distribution 
of the three cases with the real UCA ($\unit[60]{mm} \times \unit[60]{mm}$) with a 
height $r_0$ of $\unit[40]{mm}$ and with an angle $\vartheta$
from $-90$$^\circ$ till $90$$^\circ$ are presented.
One can observe that the helical phase distribution of the conventional reflector is distorted.   
The side lobes are very clear due to the existence of the antenna array in front of the reflector. 
After the reflection with the reflectors, the vortex waves 
are still achieved, but with the opposite mode, namely from mode $-1$ to mode $1$.
Fig. \ref{fig:allemode_real} shows the radiation pattern of the mode orders 
$0$, $-1$ and $-2$ with a height of $\unit[90]{mm}$ and an angle of $45$$^\circ$. 
In the case of mode order $0$, the tailored reflector is increasing the
gain from $\unit[13.9]{dBi}$ to $\unit[19.6]{dBi}$, unlike the conventional
reflector, that is decreasing the gain till $\unit[10.3]{dBi}$.
This is similar to the tailored lens. 
This is obviously an additional advantage for the tailored reflector compared to
the conventional reflector. 
For the mode order $-1$, the gain of the conventional and 
tailored reflector is increased from about $\unit[9.5]{dBi}$ 
till $\unit[16.5]{dBi}$ and $\unit[17.8]{dBi}$, respectively. The tailored reflector has $\unit[7.3]{dBi}$ 
and $\unit[1.3]{dBi}$ more than 
the UCA without reflector and the conventional reflector, respectively.
For the mode order $-2$, the gain is 
increased from $\unit[6.3]{dBi}$ till $\unit[10.4]{dBi}$ and $\unit[13.9]{dBi}$
for the conventional reflector and the tailored reflector, respectively.
The reason of the lower gain in the case of mode $-2$ 
is due to the higher beam divergence than the mode $-1$.
Same as the tailored lens, the tailored reflector has also an advantage of
saving weight and material compared to
the conventional reflector. This is when higher mode order shall be used, which can need higher distance between
the adjucent antennas. Moreover, when the number of antennas shall be incread for the utilization
of several mode orders. Please note that we are limited with manufacturing and with the equipments that we have,
therefore we didn't make a larger tailored reflector, where it is easier to observe the benefits of the tailored reflector 
compared to the conventional reflector.

\section{~reflector fabrication and measurements}

\begin{figure}[H]
    \centering
    \includegraphics[width=63mm]{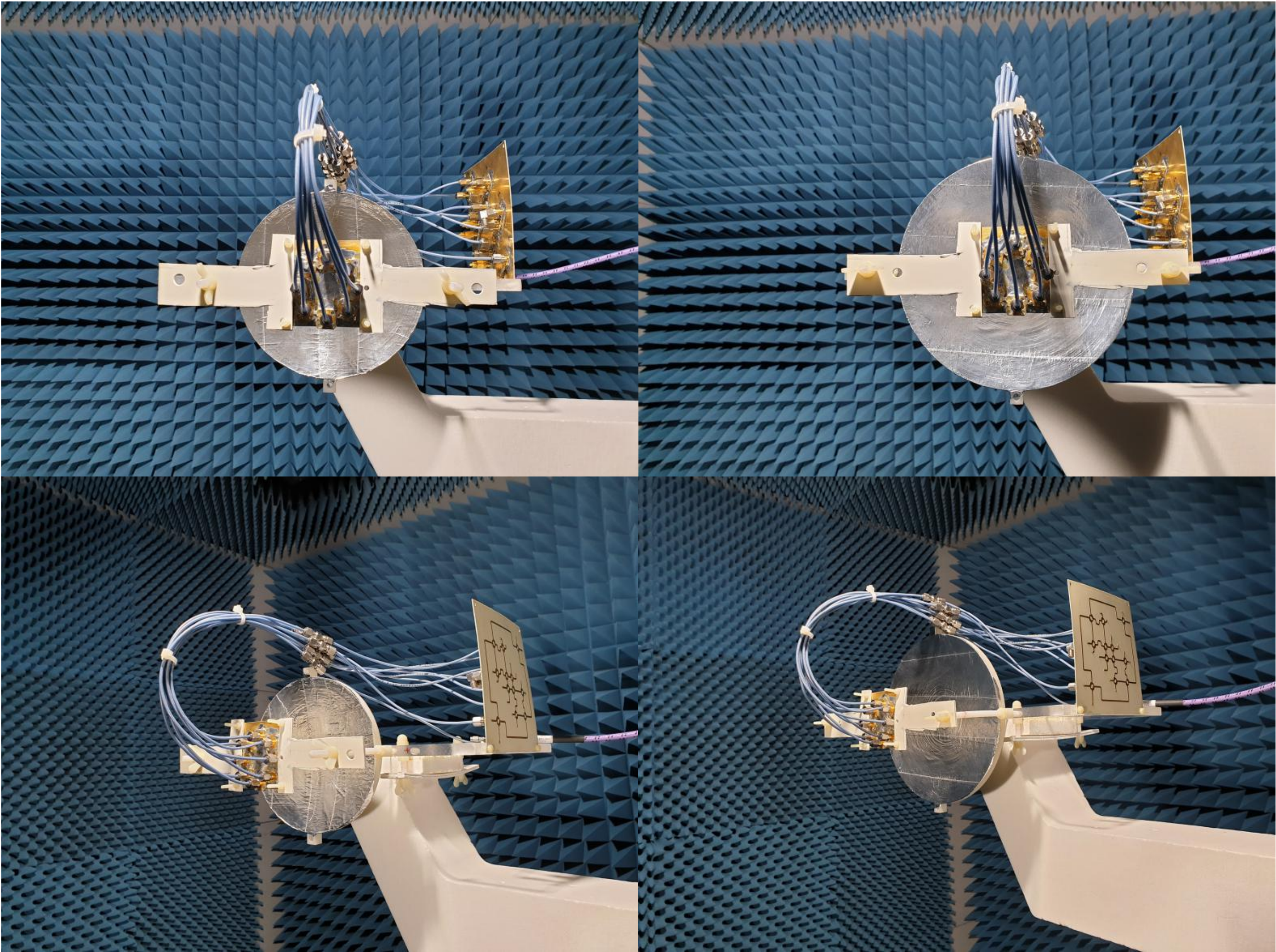}
    \caption{The manufactured conventional reflector (a,c), and tailored reflector (b,d).}
    \label{fig:Refl_manufactured}
\end{figure}

\begin{figure}[H]
    \centering
    \includegraphics[width=60mm]{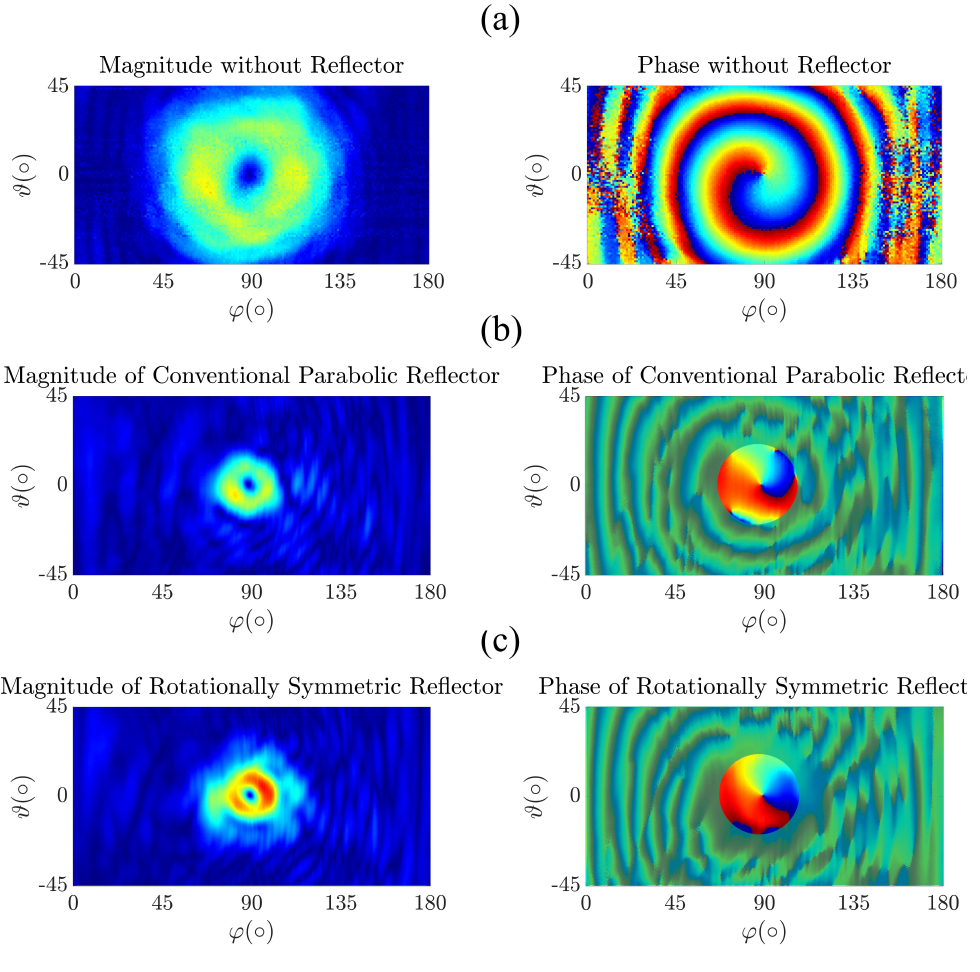}
    \caption{The amplitude and the phase distribution of antennas for the mode order $1$ without reflector (a),
    with conventional reflector (b), and with tailored reflector (c) 
    with rectangular shaped PCB $\unit[60]{mm} \times \unit[60]{mm}$.}
    \label{fig:Refl_Messung}
\end{figure}

\begin{figure}[H]
    \centering
    \includegraphics[width=80mm]{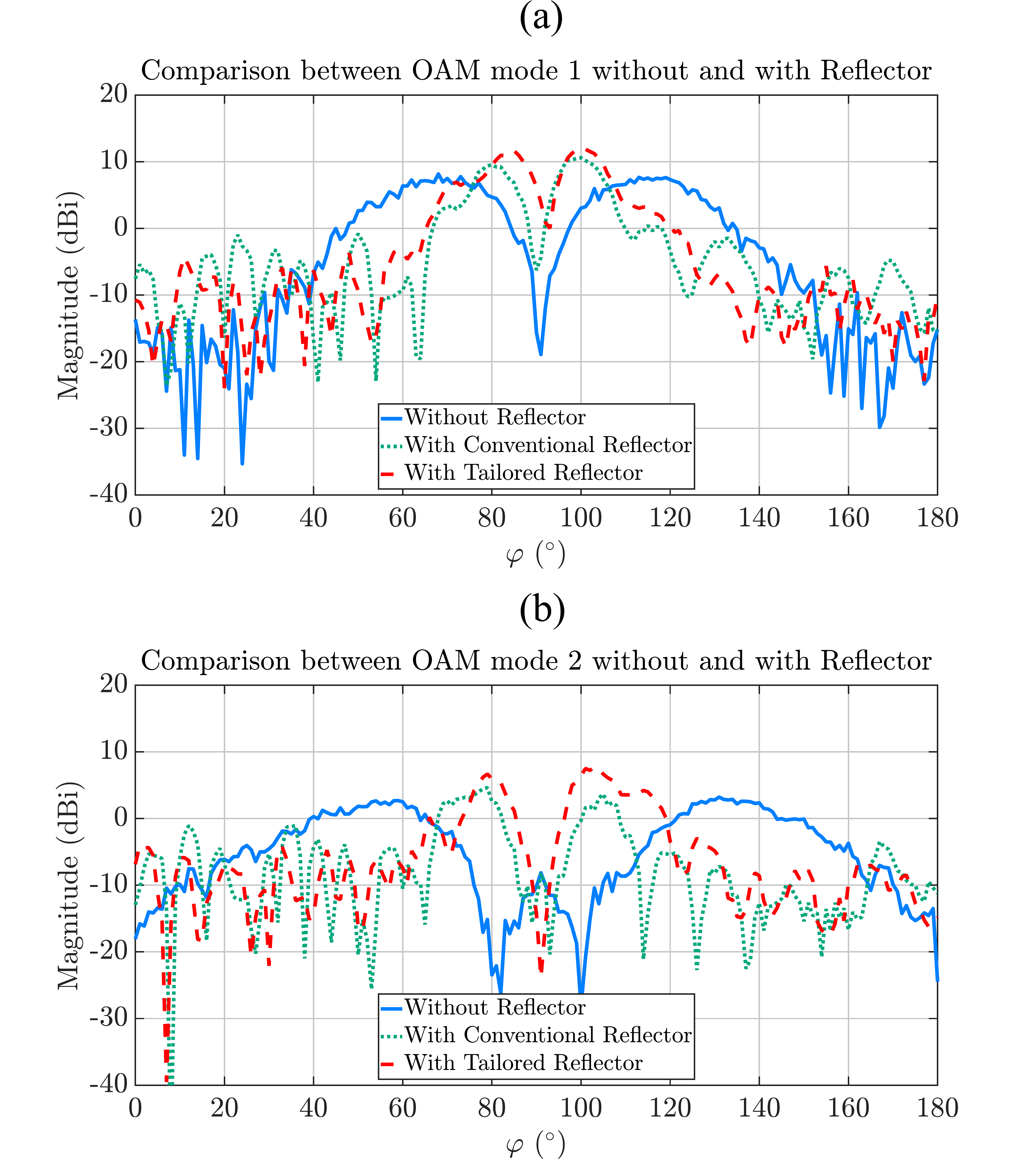}
    \caption{Comparison between UCA without reflector, with conventional reflector, 
    and with tailored reflector with rectangular shaped PCB 
    $\unit[60]{mm} \times \unit[60]{mm}$ for the mode order $1$ (a), and $2$ (b).}
    \label{fig:Refl_Messung_2}
\end{figure}

In the same way to the lens, a conventional reflector and a tailored reflector
are manufactured with a height of $\unit[90]{mm}$ and an opening angle $\vartheta$ of $45$$^\circ$
to be measured in the anechoic chamber on a rotary tabel.
The reflectors are also manufactured with polypropylene, which has to be  
covered with aluminium foil in order to reflect the incoming waves.
The reflector is assembled with UCA
and with BM.
(cf. Fig \ref{fig:Refl_manufactured}). 
In Fig. \ref{fig:Refl_Messung} the beam divergence 
of the mode order $1$ is obviously reduced. In addition, the gain of the tailored
reflector has $\unit[3.9]{dB}$ more than the antennas without reflector 
(in contrast to the simulated results) 
and $\unit[2.5]{dB}$ more than the conventional reflector (approximately same as the simulated results).
The same figure shows also the phase distribution of the three cases, which is a helical phase 
distribution of the first OAM mode order.
Please note that the phase distribution, where the amplitude 
is very high, is what we are aiming for.
Fig. \ref{fig:Refl_Messung_2} shows the magnitude of the mode order $1$ and $2$.
The mode order $2$ has a gain enhancement of $\unit[4.5]{dB}$ and $\unit[1.4]{dB}$ compared to the UCA without reflector and to the 
UCA with conventional reflector, respectively.
Please note that the measurement has many issues than the simulation due to reasons.
On one hand, the reflectors are not $100$\% smooth.
On the other hand, the aluminium layer are not perfectly glued.
Moreover, the antennas are fixed with
a piece of plastic, which will cause some absorption and delay of the waves.
Thus, the antennas and the reflector are not perfectly aligned.
Finally, the antennas are fed with eight coaxial cables, which may interfere with the path of the waves.
In order to avoid such an issue, one can integrate the BM or the power divider with
the board of the antennas, otherwise one can use two reflectors such as cassegrain reflectors.

\section*{Conclusion}

In this paper, a novel lens and reflector (tailored) are 
designed for OAM waves in order to overcome the large beam divergence 
inherent to OAM waves generated by uniform circular patch array UCA. 
The lens and the reflector are compared to
the conventional lens and to the conventional reflector.
The simulations and the measurements show that the tailored lens and the tailored
reflector have a better performance than the conventional one.
The tailored lens with a $r_0$ of $\unit[93]{mm}$ has a gain enhancement of $\unit[5.8]{dB}$ (simulated)
and $\unit[4.8]{dB}$ (measured) compared to the gain of UCA without lens and for
the first mode order. This gain enhancement is better than the conventional lens, which shows
an improvement of only $\unit[1.8]{dB}$ (simulated) and $\unit[1.7]{dB}$ (measured).
On the other hand, the tailored reflector with a 
height $r_0$ of $\unit[90]{mm}$ and an angle $\vartheta$ of 45$^\circ$ and 
with UCA (rectangular shaped PCB $\unit[60]{mm} \times \unit[60]{mm}$)
has a gain enhancement of $\unit[8.3]{dBi}$ (simulated) and $\unit[3.9]{dB}$ (measured)
compared to the gain of UCA without reflector and for the first mode order.
The conventional lens has less gain enhancement of only $\unit[7]{dB}$ (simulated) and $\unit[2.5]{dB}$ (measured).
In addition, the results of the tailored reflector show that 
the tailored reflector is more efficient than the conventional reflector 
until it reaches a height $r_0$ of $1.5$$\lambda$ and an angle $\vartheta$ of $38$$^\circ$, separately.
Moreover, the tailored lens and reflector have also additional advantage.
This advantage appears when higher mode order shall be used, which need 
a higher distance between the adjucent antennas in order to get the maximum gain with less sie lobes.
Furthermore, when the number of antennas shall be incread for the utilization
of several mode orders in the case of OAM target localization.
This two tailored component can save weight and material compared to
the conventional lens and reflector. 
In conclusion, the lens has a big advantage for clear results compared to the reflector.
The size of the antennas is not a problem for the lens, in contrast to the
reflector where the vortex waves can be disturbed by the UCA and the cables, which are on the way
of the waves. Nevertheless, the lens consume too much material, which leads to limit the gain 
enhancement.




\section*{Acknowledgment}

This work was funded by the Deutsche Forschungs-
gemeinschaft (DFG, German Research Foundation) – Project-ID $287022738$ – 
TRR $196$ in the framework of projects S$03$, and M$02$.

\bibliography{OAM_Reflector_lens_ARxiv}
\bibliographystyle{IEEEtran}

\section*{Biographies}


\begin{wrapfigure}{l}{25mm} 
\includegraphics[width=1in,height=1.25in,keepaspectratio]{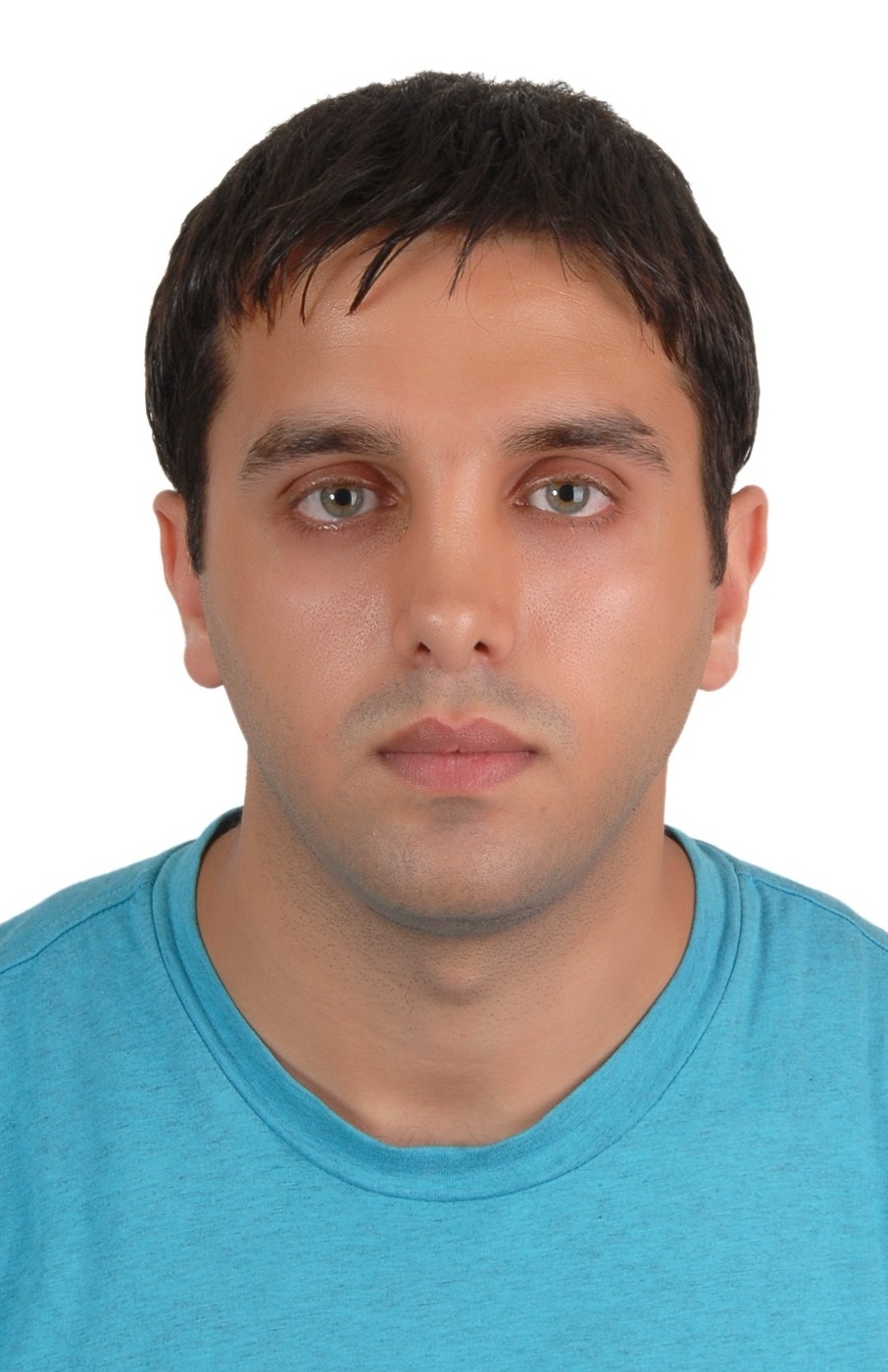}
\end{wrapfigure}
\textbf{Mohamed Haj Hassan}
was born in Beirut,
Lebanon. He received the B.Sc. and M.sc. degree in Electrical
Engineering/High-Frequency Technology from the Technical University of Berlin, Berlin, Germany, in $2010$ and $2012$,
respectively. From $2015$ to $2017$ he was working at the Technical University of Ilmenau, Ilmenau, Germany,
in the field of Ground Penetration Radar GPR.
Since $2017$ he is a member of the Laboratory of General and Theoretical Electrical 
Engineering of the University of Duisburg-Essen.
His research interests include RF and antenna technology, mm-waves, vortex waves,
electromagnetic metamaterials, and computational electromagnetics. 
\vfill

\begin{wrapfigure}{l}{25mm} 
    \includegraphics[width=1in,height=1.25in,keepaspectratio]{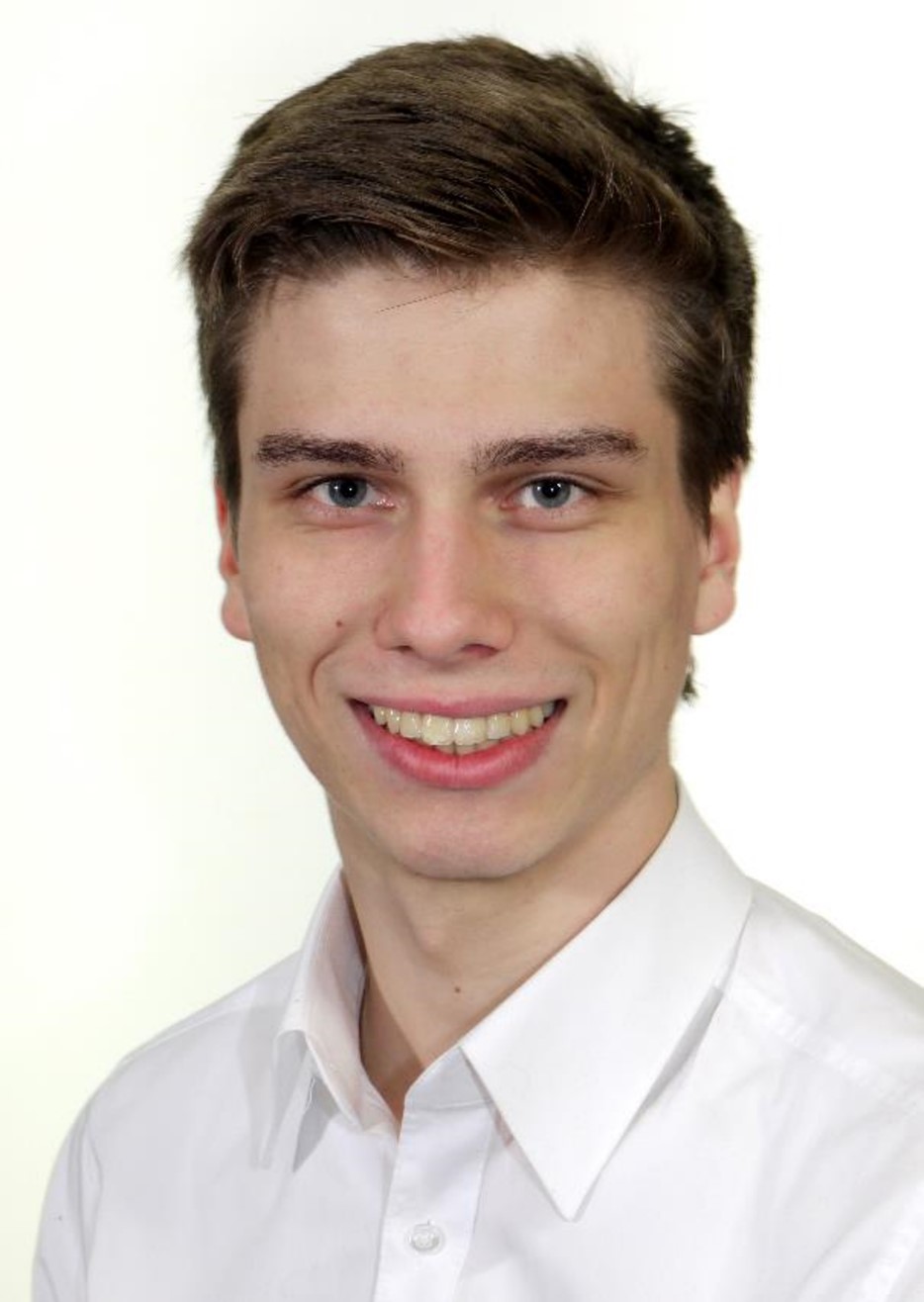}
    \end{wrapfigure}
    \textbf{Benedikt Sievert}
    was born in Krefeld, Germany. He received his B.Sc. and M.Sc. in Electrical 
    Engineering/High-Frequency Systems from the University of Duisburg-Essen in 
    $2017$ and $2019$, respectively. 
    Since $2017$ he is a member of the Laboratory of General and Theoretical Electrical 
    Engineering of the University of Duisburg-Essen. His research interests include 
    mm-wave on-chip antennas, electromagnetic metamaterials, theoretical and computational electromagnetics. 
    \vfill

\begin{wrapfigure}{l}{25mm} 
\includegraphics[width=1in,height=1.25in,keepaspectratio]{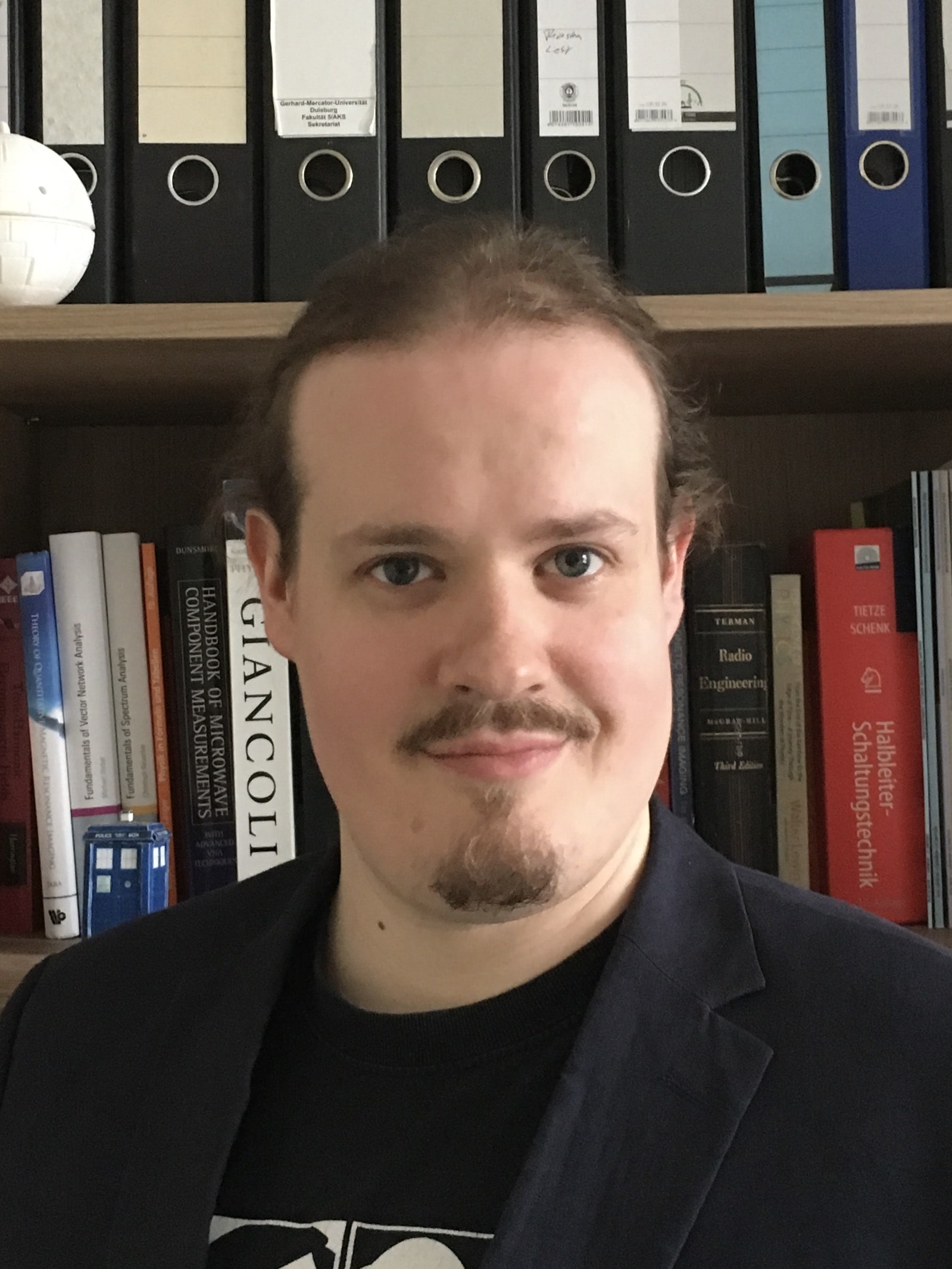}
\end{wrapfigure}
\textbf{Jan Taro Svejda}
    started his electrical engineering career at the University of 
    Applied Science, Düsseldorf, Germany, where he received his B.Sc. 
    degree in $2008$. Consecutively he continued his studies in Electrical 
    Engineering and Information Technology at the University of Duisburg-Essen, 
    Duisburg, Germany, and received his M.Sc. degree in $2013$ and his Dr.-Ing. Degree 
    in $2019$ for his research work in the field of X-nuclei based magnetic 
    resonance imaging, respectively. He is currently working as a research 
    assistant at University of Duisburg-Essen in the department of General 
    and Theoretical Electrical Engineering where he is involved in teaching 
    several lectures and courses mainly in the field of electrical engineering.
    His general research interest includes all aspects of theoretical and applied 
    electromagnetics, currently focusing on medical applications, electromagnetic 
 metamaterials, and scientific computing methods. 
\vfill

\begin{wrapfigure}{l}{25mm} 
\includegraphics[width=1in,height=1.25in,keepaspectratio]{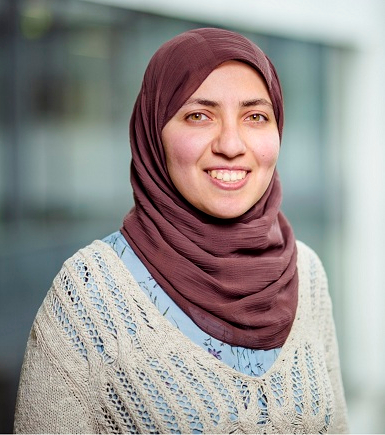}
\end{wrapfigure}
\textbf{Aya Mostafa Ahmad}
received the B.Sc. and M.S.C degrees in electrical engineering from 
the German University in Cairo, Egypt, in $2011$ and $2014$ respectively.
She is currently pursuing the Ph.D. degree with the Institute of Digital Communication 
Systems, Ruhr-Universität Bochum, Germany. Her research interests include MIMO radar 
signal processing, waveform design optimization, cognitive radars, direction of 
arrival (DOA) algorithms and machine learning applications for radar resources management.
\vfill

\begin{wrapfigure}{l}{25mm} 
\includegraphics[width=1in,height=1.25in,keepaspectratio]{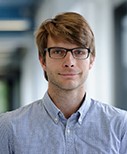}
\end{wrapfigure}
\textbf{Jan Barowski}
received the B. Sc. and M. Sc. in electrical engineering 
from Ruhr-University Bochum, Bochum, Germany, in $2010$ and $2012$,
respectively. Since $2012$ he is with the Institute of Microwave 
Systems, headed by Ilona Rolfes, Ruhr-University Bochum, as a 
Research Assistant. In $2017$ he received the Dr.-Ing. degree in 
electrical engineering from Ruhr-University Bochum and is now 
working as post-doctoral Research Scientist at the Institute 
of Microwave Systems. His currents fields of research are concerned
with radar signal processing, radar imaging and material characterization techniques. 
\vfill

\begin{wrapfigure}{l}{25mm} 
    \includegraphics[width=1in,height=1.25in,keepaspectratio]{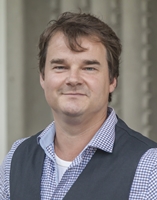}
    \end{wrapfigure}
    \textbf{Andreas Rennings}
    studied electrical engineering at the University of Duisburg-Essen,
    Germany. He carried out his diploma work during a stay at the University 
    of California in Los Angeles. He received his Dipl.-Ing. and Dr.-Ing. 
    degrees from the University of Duisburg-Essen in $2000$ and $2008$, respectively.
    From $2006$ to $2008$ he was with IMST GmbH in Kamp-Lintfort, Germany, where 
    he worked as an RF engineer. Since then, he is a senior scientist and 
    principal investigator at the Laboratory for General and Theoretical 
    Electrical Engineering of the University of Duisburg-Essen. His 
    general research interests include all aspects of theoretical and 
    applied electromagnetics, currently with a focus on medical 
    applications and on-chip millimeter-wave/THz antennas. He received 
    several awards, including a student paper price at the $2005$ IEEE Antennas 
    and Propagation Society International Symposium and the VDE-Promotionspreis $2009$ for the dissertation. 
\vfill

\begin{wrapfigure}{l}{25mm} 
    \includegraphics[width=1in,height=1.25in,keepaspectratio]{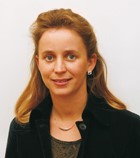}
    \end{wrapfigure}
    \textbf{Ilona Rolfes}
    (Member, IEEE) received the Dipl.-Ing. and 
    Dr.-Ing. degrees in electrical engineering from Ruhr University 
    Bochum, Bochum, Germany, in $1997$ and $2002$, respectively. From $1997$
    to $2005$, she was with the High Frequency Measurements Research 
    Group, Ruhr University Bochum, as a Research Assistant. From $2005$ 
    to $2009$, she was a Junior Professor with the Department of 
    Electrical Engineering, Leibniz University Hannover, Hannover, 
    Germany, where she became the Head of the Institute of 
    Radio frequency and Microwave Engineering in $2006$. Since 
    $2010$, she has been leading the Institute of Microwave Systems,
    Ruhr University Bochum. Her fields of research concern high-frequency 
    measurement methods forvector network analysis, material characterization,
    noise characterization of microwave devices, sensor principles for radar 
    systems, and wireless solutions for communication systems.
\vfill

\begin{wrapfigure}{l}{25mm} 
    \includegraphics[width=1in,height=1.25in,keepaspectratio]{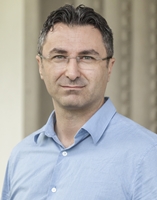}
    \end{wrapfigure}
    \textbf{Aydin Sezgin}
    received the Dipl.Ing. (M.S.) degree in communications engineering from Technische
Fachhochschule Berlin (TFH), Berlin, in $2000$, and the Dr. Ing. (Ph.D.) degree in 
electrical engineering from TU Berlin, in $2005$. From $2001$ to $2006$, he was with 
the Heinrich-Hertz-Institut, Berlin. From $2006$ to $2008$, he held a Post-doctoral 
position, and was also a Lecturer with theInformation Systems Laboratory, 
Department of Electrical Engineering, Stanford University, Stan-ford, CA, USA. 
From $2008$ to $2009$, he held a Postdoctoral position with the Department of 
Electrical Engineering and Computer Science, Universityof California, Irvine, 
CA, USA. From $2009$ to $2011$, he was the Head of the Emmy-Noether Research Group 
on Wireless Networks, Ulm University. In $2011$, he joined TU Darmstadt, Germany, 
as a Professor. He is currently a Professor of information systems and sciences 
with the Department of Electrical Engineering and Information Technology, 
Ruhr-Universität Bochum, Germany. His research interests include in signal 
processing, communication, and information theory, with a focus on wireless 
networks. He has published several book chapters more than $40$ journals and $140$ 
conference papers in these topics. He has coauthored a book on multiway 
communications. He is a winner of the ITG-Sponsorship Award, in $2006$. He 
was a first recipient of the prestigious Emmy-Noether Grant by the German 
Research Foundation in communication engineering, in $2009$. He has coauthored papers 
that received the Best Poster Award at the IEEE Communication Theory Workshop, 
in $2011$, the Best Paper Award at ICCSPA, in $2015$, and the Best Paper Award at 
ICC, in $2019$. He has served as an Associate Editor for the IEEE T\textsubscript{RANSACTIONS} 
\textsubscript{ON} W\textsubscript{IRELESS} C\textsubscript{OMMUNICATIONS}, from $2009$ to $2014$.
\vfill

\begin{wrapfigure}{l}{25mm} 
    \includegraphics[width=1in,height=1.25in,keepaspectratio]{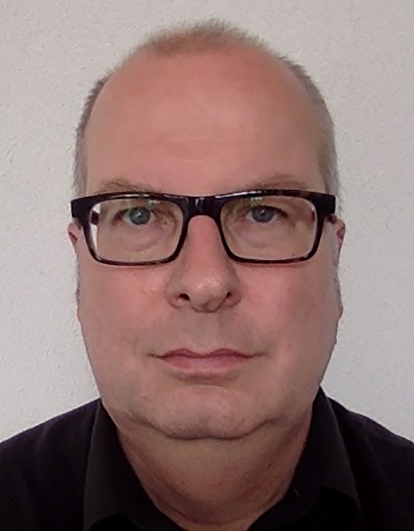}
    \end{wrapfigure}
    \textbf{Daniel Erni}
    received a diploma degree from the University of Applied Sciences in 
    Rapperswil (HSR) in $1986$, and a diploma degree from ETH Zürich in $1990$,
    both in electrical engineering. Since $1990$ he has been working at the 
    Laboratory for Electromagnetic Fields and Microwave Electronics, ETH 
    Zürich, where he got his Ph.D. degree in laser physics $1996$. From $1995$-$2006$ 
    he has been the founder and head of the Communication Photonics Group at ETH 
    Zürich. Since Oct. $2006$ he is a full professor for General and Theoretical
    Electrical Engineering at the University of Duisburg-Essen, Germany. 
    His current research interests include optical interconnects, 
    nanophotonics, plasmonics, advanced solar cell concepts, optical 
    and electromagnetic metamaterials, RF, mm-wave and THz engineering, 
    biomedical engineering, bioelectromagnetics, marine electromagnetics, 
    computational electromagnetics, multiscale and multiphysics modeling, 
    numerical structural optimization, and science and technology studies 
    (STS). Daniel Erni is a co-founder of the spin-off company airCode on 
    flexible printed RFID technology. He is a Fellow of the Electromagnetics 
    Academy, a member of the Center for Nanointegration Duisburg-Essen (CeNIDE),
    as well as a member of the Swiss Physical Society (SPS), of the German 
    Physical Society (DPG),of the Optical Society of America (OSA), and of the IEEE. 
\vfill


\end{document}